\documentclass[sigconf,nonacm,screen]{acmart}
\AtBeginDocument{%
  }

\usepackage{appendix}

\usepackage{graphicx}
\usepackage{microtype}
\usepackage{xspace}
\usepackage{subcaption}
\usepackage{wrapfig}
\usepackage{algorithm}
\usepackage{amsthm}
\usepackage{verbatim}
\usepackage{mathtools}
\usepackage{soul}
\usepackage{enumitem}
\usepackage{bbm}
\usepackage{multirow}
\usepackage{listings}
\usepackage{adjustbox}

\newcommand{\tsd}{360$^\circ$ }

\setcopyright{none}               % No ACM copyright

\renewcommand\footnotetextcopyrightpermission[1]{}  % Remove footnote on first page
\usepackage{titlesec}
\usepackage{lipsum}% just to generate text for the example

\titlespacing*{\section}
{0pt}{1ex}{1ex}
\titlespacing*{\subsection}
{0pt}{1ex}{1ex}
\titlespacing*{\subsubsection}
{0pt}{1ex}{1ex}

\begin{document}

\title{Securing Immersive 360 Video Streams through Attribute-Based Selective Encryption}

\author{Mohammad Waquas Usmani}
\email{mohammadwaqu@umass.edu}
\affiliation{%
  \institution{University of Massachusetts Amherst}
  % \city{Amherst}
  \state{Massachusetts}
  \country{USA}
}
\author{Susmit Shannigrahi}
\email{sshannigrahi@tntech.edu}
\affiliation{%
  \institution{Tennessee Technological University}
  % \city{Cookeville}
  \state{Tennessee}
  \country{USA}
}
\author{Michael Zink}
\email{zink@ecs.umass.edu}
\affiliation{%
  \institution{University of Massachusetts Amherst}
  % \city{Amherst}
  \state{Massachusetts}
  \country{USA}
}

\begin{abstract}
Delivering high-quality, secure \tsd video content introduces unique challenges, primarily due to the high bitrates and interactive demands of immersive media. Traditional HTTPS-based methods, although widely used, face limitations in computational efficiency and scalability when securing these high-resolution streams. To address these issues, this paper proposes a novel framework integrating Attribute-Based Encryption (ABE) with selective encryption techniques tailored specifically for tiled \tsd video streaming. Our approach employs selective encryption of frames at varying levels to reduce computational overhead while ensuring robust protection against unauthorized access.

Moreover, we explore viewport-adaptive encryption, dynamically encrypting more frames within tiles occupying larger portions of the viewer’s field of view. This targeted method significantly enhances security in critical viewing areas without unnecessary overhead in peripheral regions. We deploy and evaluate our proposed approach using the CloudLab testbed, comparing its performance against traditional HTTPS streaming. Experimental results demonstrate that our ABE-based model achieves reduced computational load on intermediate caches, improves cache hit rates, and maintains comparable visual quality to HTTPS, as assessed by Video Multimethod Assessment Fusion (VMAF).

\vspace{0.3em}
\end{abstract}

\maketitle

\vspace{-1em}
\section{Introduction} \label{sec:introduction}

Digital Rights Management (DRM) has played a pivotal role in the widespread adoption and success of online video streaming by protecting content from unauthorized access and distribution. Traditional DRM approaches secure video content in transit and at rest separately, primarily relying on Hypertext Transfer Protocol Secure (HTTPS) coupled with Transport Layer Security (TLS).
The success of video streaming is also based on the support by Content Distribution Networks (CDN)~\cite{maggs2015algorithmic,Adhikari2015Netflix}, which make streaming scalable to hundreds of millions of users.

However, employing HTTPS with adaptive bitrate (ABR) streaming—commonly implemented through protocols such as Dynamic Adaptive Streaming over HTTP (DASH) on top of CDN distribution architectures—introduces significant computational overhead due to the necessity of decrypting and re-encrypting content at intermediate caching nodes. This limitation becomes particularly pronounced in bandwidth-intensive scenarios such as \tsd video streaming, where the interactive nature and high bitrate require additional networking and computing resources.

To overcome these limitations, prior work~\cite{10.1145/3712676.3714450} introduced Attribute-Based Encryption for DRM-enabled conventional video streaming. This approach secures the data itself, removing the need for separate transport-layer encryption. It simplifies content distribution by encrypting data once with specific attributes, allowing caching nodes to serve encrypted video without additional cryptographic operations. Clients obtain attribute-based keys from a license server to decrypt and access the content, significantly reducing the computational load on caches and facilitating scalable streaming.

The work presented in this paper addresses the challenges of securely and efficiently streaming \tsd videos. Due to their spherical nature, \tsd videos are projected onto two-dimensional planes, tiled, and selectively streamed based on the user's viewport~\cite{8643410}. Depending on the viewer's head movement, their viewport might be composed of portions of different tiles. Consequently, the video frames of several tiles have to be streamed to the client in parallel, which is in contrast to conventional video, where there is only a maximum of one stream. Our approach introduces selective frame encryption strategies tailored explicitly to the tiled structure of \tsd videos. 
This 2-dimensional method combines frame-selective and tile-selective encryption to reduce computational overhead while preserving content protection. Different frame types are encrypted at varying levels based on each tile’s relevance to the viewport.

% Our approach is two-dimensional, by combining a frame-selective approach (1st dimension) with a tile-selective approach (2nd dimension). We reduce computational overhead without compromising content protection by adaptively encrypting different frame types at varying security levels, depending on the tile's contribution to the viewport.

We evaluate our proposed approach in CloudLab~\cite{duplyakin2019cloudlab}, comparing it directly against conventional HTTPS-based streaming. Experimental results indicate that employing ABE substantially reduces computational overhead at intermediate caches, while preserving cache efficiency and maintaining comparable video quality, as assessed by the Video Multimethod Assessment Fusion (VMAF) metric.
% Our experiments demonstrate that using ABE significantly reduces computational demands at intermediate caches while maintaining high cache efficiency and delivering comparable video quality, as measured by Video Multimethod Assessment Fusion (VMAF). 
This work illustrates the benefits of integrating ABE into tiled \tsd video streaming, and also establishes a promising pathway toward scalable and efficient secure immersive media distribution.

This paper makes the following contributions. We introduce an ABE-based two-dimensional selective encryption method, which is specifically designed for scalable streaming of \tsd video. In addition, we conduct an in-depth evaluation, with two distinct experiment setups: a small-scale experimental and a large-scale multicache hierarchical experiment with strict bandwidth limitations. 
The results of this evaluation show that CPU usage at both the server and caches is significantly reduced with up to 63\% reduction at the cache. In several cases, it also delivers improved hit rates, while maintaining QoE comparable to the TLS-based approach.

% \vspace{-0.8em}
\section{Background and Related Work} \label{sec:background}

\subsection{Fundamentals of \tsd Video}
\begin{figure}[h]
% \centering
% \begin{minipage}{0.2\textwidth}
% \centering
% \includegraphics[width=.9\linewidth]{figs/3degreefreedom.png}
% \captionsetup{justification=centering}
% \caption{3 Degrees of Freedom}
% \label{fig:3degree}
% \end{minipage}%
% \hfill
% \begin{minipage}{0.8\textwidth}
\centering
\includegraphics[width=.9\linewidth]{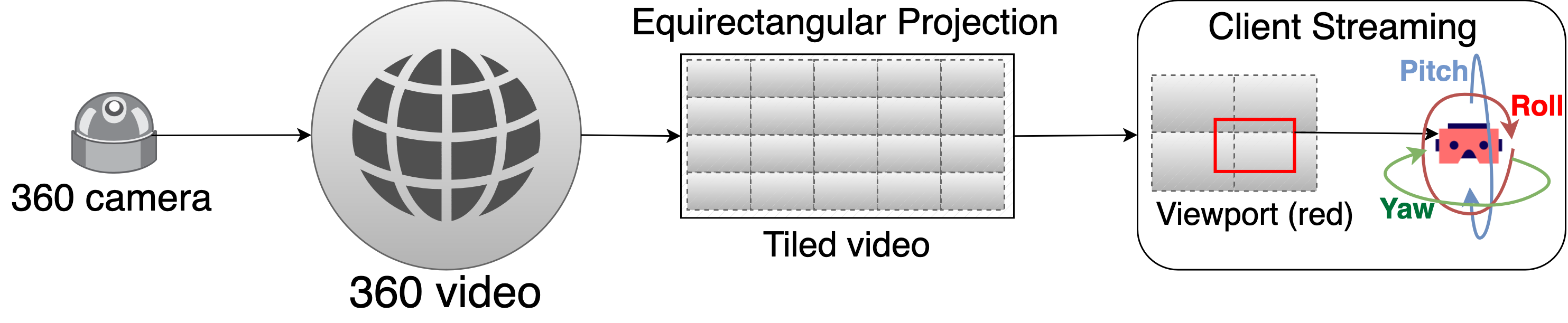}
\captionsetup{justification=centering}
\caption{\tsd Video Streaming Process}
\label{fig:360stream}
% \end{minipage}
\end{figure}

\tsd videos offer immersive experiences by capturing a complete spherical view, enabling viewers to interactively explore the environment.
 As illustrated in Fig.~\ref{fig:360stream}, these videos provide three degrees of freedom (3DoF): yaw (left/right), pitch (up/down), and roll (rotational movement), contributing to an engaging user experience.

\textbf{Equirectangular Projection:} To encode spherical video data using standard video codecs, a spherical view must be projected onto a two-dimensional plane. The most widely used method is equirectangular projection~\cite{Graf:2017:Equi,Zare:2016:Equi}, mapping the spherical surface onto a rectangular grid. This allows compatibility with conventional video formats and codecs while preserving spatial relationships crucial for accurate spherical rendering during playback.

\textbf{Viewport:} A viewport represents the visible portion of the spherical content, typically defined by a horizontal and vertical field of view (FOV) ranging from 90 to 120 degrees. The viewport dynamically adjusts based on user interactions via VR headsets, mobile devices, or computer interfaces. By streaming primarily the viewport region, substantial bandwidth and computational savings are achieved without diminishing the immersive experience~\cite{Graf:2017:Equi}.

\textbf{Tiling:} Delivering high-quality streaming across an entire \tsd video simultaneously is bandwidth-intensive and inefficient. To optimize performance, videos are spatially divided into smaller rectangular sections known as tiles. These tiles, derived from the equirectangular projection, allow adaptive streaming by enabling higher-quality delivery for tiles within the user's viewport, while peripheral tiles are streamed at lower quality or omitted. This technique significantly reduces bandwidth consumption and ensures smooth playback even under constrained network conditions~\cite{Zeynali:2024:BOLA360}.

Fig.~\ref{fig:360stream} outlines the complete streaming workflow: capturing spherical video content, projecting it onto an equirectangular plane, partitioning it into tiles, and selectively streaming high-quality tiles based on the user's real-time viewport.

\subsection{Attribute-Based Encryption (ABE)} ABE is a cryptographic technique designed to enable fine-grained access control for encrypted data. Unlike traditional symmetric encryption methods like AES, 
%which rely on a single key for both encryption and decryption, 
ABE uses attributes—descriptive elements assigned to users or data—to define access policies. In this scheme, data can be decrypted only if a user’s attributes satisfy the access policy associated with the encrypted content.

The operation of ABE begins with a setup phase conducted by a Trusted Authority (TA). During this phase, the TA generates a Master Key and a Public Key. The Data Owner encrypts content using the Public Key and defines an access policy based on attributes. The TA, using the Master Key, generates Private Keys for users, tailored to their specific attributes. When a user attempts to decrypt a ciphertext, the system verifies whether their attributes satisfy the encryption policy. If the attributes match, the decryption succeeds, granting access to the content. In our work, we utilized the \textbf{CPABE-Toolkit}, a command-line tool \cite{CPABE}, which is based on the cryptographic framework outlined in \cite{4223236}. Prior work \cite{10.1145/3712676.3714450} has utilized ABE to provide segment-level security for conventional video. In this work, the authors apply ABE to individual video frames within each segment.

\subsection{Selective Encryption of Video Streams}  Selective encryption has been a widely researched topic in multimedia\cite{yeung2009partial,DBLP:journals/corr/abs-2201-03391,10.5555/968883.969421}. A recent study~\cite{DBLP:journals/corr/abs-2201-03391}, explores the use of Advanced Encryption Standard (AES) for selectively encrypting H264/AVC videos. This system processes the H264 bitstream to identify \textit{I}-frames and encrypts them using AES, based on the assumption that removing access to \textit{I}-frames renders the dependent \textit{P} and \textit{B}-frames ineffective. While it is true that \textit{I}-frames are self-contained and provide all the necessary information to display a complete image \cite{Salomon_2007}, this approach overlooks the fact that \textit{P} and \textit{B}-frames may still carry meaningful information. \textit{P} and \textit{B}-frames contain \textit{I}-blocks, and when a sequence of these frames is correlated with their reference frames, they can still convey significant visual information \cite{492420}. Building on this observation, our work explores encrypting \textit{P} and \textit{B}-frames in addition to \textit{I}-frames in H264 bitstreams, employing varying levels of encryption. This approach aims to address the limitations of solely encrypting \textit{I}-frames by targeting the residual visual information carried by dependent frames, thereby enhancing security and reducing the risk of meaningful content leakage.

%Several other selective encryption methods have been explored in the literature, including encrypting transformation matrix \cite{yeung2009partial} and motion vectors \cite{10.5555/968883.969421}. Investigating the effects of ABE in conjunction with these techniques could yield interesting insights and presents an avenue for future research.

\subsection{VMAF for \tsd Video QOE Evaluation} Video Multi-method Assessment Fusion (VMAF)~\cite{vmaf} is a video quality metric that combines human visual perception models with machine learning techniques to evaluate video quality. It assesses how closely a distorted or degraded video resembles a reference video, providing an objective measure of user experience. VMAF scores range from 0 to 100, with higher scores indicating better video quality and a more enjoyable viewing experience.

The metric is widely used in video streaming and encoding research to quantify quality degradation and optimize video delivery systems. Specifically, Orduna et al.~\cite{8924618} explore the application of VMAF for \tsd videos. While VMAF was originally designed to evaluate conventional 2D videos, \cite{8924618} demonstrated that it can also be effectively applied to \tsd videos without any modifications. Their evaluation was based on leveraging a diverse dataset of \tsd videos in equirectangular projection format.

\subsection{HTTP-based Video Streaming and CDNs}
DASH~\cite{sodagar2011mpeg} and HLS~\cite{apple-hls} are popular ABR streaming standards that segment videos at various bitrates, allowing clients to adapt quality via algorithms like BOLA~\cite{spiteri2020bola} or Pensieve~\cite{mao2017neural}. HTTPS secures traffic via TLS~\cite{rfc8446}, but its point-to-point encryption adds computational overhead, especially when \tsd videos are cached.

% MPEG-DASH~\cite{sodagar2011mpeg} and HTTP Live Streaming (HLS)\cite{apple-hls} are widely used standards for delivering adaptive bitrate (ABR) video over HTTP. Videos are segmented into multiple bitrates, allowing clients to select optimal quality based on bandwidth using algorithms such as BOLA\cite{spiteri2020bola} or Pensieve~\cite{mao2017neural}. Typically, HTTPS secures streaming traffic through Transport Layer Security (TLS)~\cite{rfc8446}, but its point-to-point encryption increases computational load, especially when \tsd videos are cached.

Content Distribution Networks (CDNs) efficiently distribute digital content by caching it close to users, minimizing buffering and latency. Popular streaming services leverage CDNs extensively, employing caches globally to ensure high-quality service. However, the use of HTTPS complicates caching by necessitating TLS termination and re-encryption at CDN nodes, significantly adding overhead to streaming systems~\cite{10.1007/978-3-031-28486-1_27,10.1145/3712676.3714450}.

% \subsection{Caching for video delivery} For video delivery, existing caching solutions fall broadly into two categories. In the first category, the algorithms exploit the sequential access patterns of video segments to make better caching decisions. For example, CoLeap \cite{shi2020coleap} describes a pre-fetching-based cache algorithm that uses a QoE-oriented deep neural network model to decide the content to pre-fetch into the cache. Whereas \cite{10.5555/AAI28544902} proposes performing smart prefetching by forecasting throughput values in the cache using previous client throughput data, thus optimizing resource utilization by pre-fetching video segments based on predicted demand, while improving client QoE. Another approach, presented in \cite{10.1145/2713168.2713181}, introduces an integrated pre-fetching and caching proxy (ipac), a novel 0.5-competitive online pre-fetching algorithm specifically designed for HTTP-based adaptive streaming services like Netflix and YouTube, without requiring modifications to content servers or clients that improves byte-hit ratios and video rates.
% In the second category, the cache algorithms make decisions about specific qualities of each video segment to cache to maximize the cache hitrate for a single as well as a network of caches \cite{poularakis2016caching,  araldo2016representation}. 
% These approaches are orthogonal to \acro, which does not focus on admission and eviction strategies for video caching. Since \acro is agnostic to these strategies, any of them can be integrated in the streaming system we propose.

\section{System Architecture} \label{sec:architechture}

The architecture we created for ABE-based \tsd video distribution is inspired by the work presented in~\cite{10.1145/3712676.3714450}. While previous work has primarily focused on conventional video streaming, our approach is specifically tailored to \tsd video content, requiring the design of an architecture that enables selective encryption for tiled \tsd videos. Our proposed architecture is realized in a prototype implementation, which we use for the evaluation of our approach.

\begin{figure}[!ht]\centering
\includegraphics[width=0.65\columnwidth]{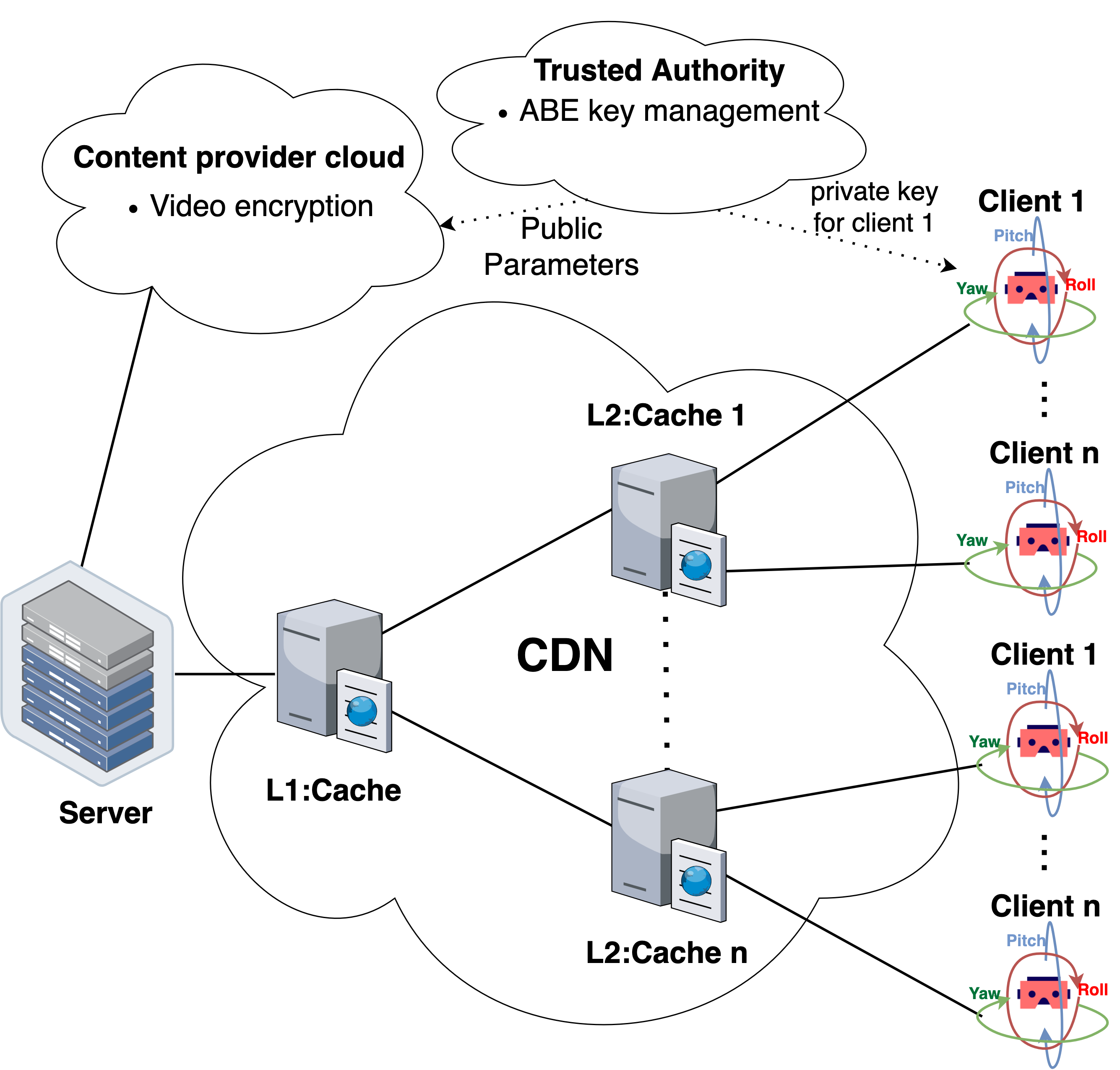}
\captionsetup{justification=centering}
\caption{System Architecture}
\label{sys-arch}
\end{figure}

\subsection{HTTP+ABE for Secured %\tsd Video 
Streaming:}

Figure~\ref{sys-arch} illustrates the system architecture of our ABE-based \tsd video streaming framework, integrated into a typical DASH-based distribution pipeline.

The architecture consists of three key entities: Content Provider, the CDN, and a Trusted Authority. It includes three main components: origin servers, CDN edge servers (caches), and clients (e.g, head-mounted devices (HMD), laptops, phones). The Content Provider divides the equirectangular projection of the \tsd videos into tiles and encodes each into multiple qualities, generates DASH metadata (e.g., MPD), and handles content encryption using ABE. All content is originally stored on the origin server.

Client requests are directed to edge servers based on CDN policies~\cite{maggs2015algorithmic}. If the requested content is cached, the edge server delivers it directly; otherwise, it retrieves it from the origin server and decides whether to cache it for future use.

\textbf{Integration into HTTP-based Streaming: }
To integrate ABE into standard DASH streaming, we made several modifications:  A Trusted Authority (TA) generates a master key and a public key. The public key is used to encrypt video segments using an ABE policy, while the TA uses the master key to generate user-specific private keys embedded with access attributes. Unlike HTTPS-based approaches, where key exchange is handled via TLS, in the case of ABE, a separate mechanism is used where the TA issues decryption keys directly to clients. CDN caches remain agnostic to the decryption process—they store encrypted segments but do not perform any cryptographic operations.

Our system also introduces selective encryption of video streams (see Sect.~\ref{subsec:selective-enc}). While prior work~\cite{QiaoEncrypt2001,ShiEncryption1998} has explored selective encryption, modern HTTP-based streaming is typically binary: either fully encrypted (via HTTPS) or not encrypted (HTTP) at all. To support selective encryption in DASH, it is required that the server or cache can communicate to the client which portions of the video stream are encrypted. In our approach, this is achieved by modifying the MPD to include ``encryption level" for a segment that indicates what video frames are encrypted. On the client side, once a segment is downloaded, the DASH player reads this encryption level and invokes selective decryption logic. The client uses its private key to decrypt the targeted frames before playback. Additional implementation details are provided in Section~\ref{subsec:protoype}.

All other components of standard DASH remain unchanged, allowing seamless integration into existing streaming infrastructures.

\subsection{Selective Encryption for Tiled \tsd Videos} \label{subsec:selective-enc}

To balance content protection and computational efficiency, we propose a flexible, frame-based, two-dimensional
selective encryption approach tailored to \tsd video streaming. Rather than striving for complete secrecy, \emph{our aim is to significantly impair unauthorized viewing by degrading video quality through partial encryption}. This strategy represents a trade-off between the extent of encryption and the resources consumed during en- and decryption. In addition, \emph{our encryption approach is specifically designed for tile-based \tsd streaming through the introduction of a scheme that adjusts the level of encryption based on the importance of a tile}.

In the first dimension, our method supports multiple levels of encryption, ranging from encrypting only \textit{I-}frames to encrypting \textit{P}-frames or all frames, providing scalability in protection and performance. Designed for H264/AVC encoded videos, we parse Network Abstraction Layer (NAL) units from DASH segments to identify \textit{I, P, }and \textit{B}-frames. Selected frames are encrypted using the CPABE toolkit with attribute-based policies, and reinserted into their original locations. We update frame size headers accordingly but intentionally avoid modifying the MP4 container’s metadata and offsets, rendering the video unplayable unless decrypted correctly, adding an additional layer of access control.

\begin{figure}[h]\centering
  \includegraphics[width=1.05\linewidth]{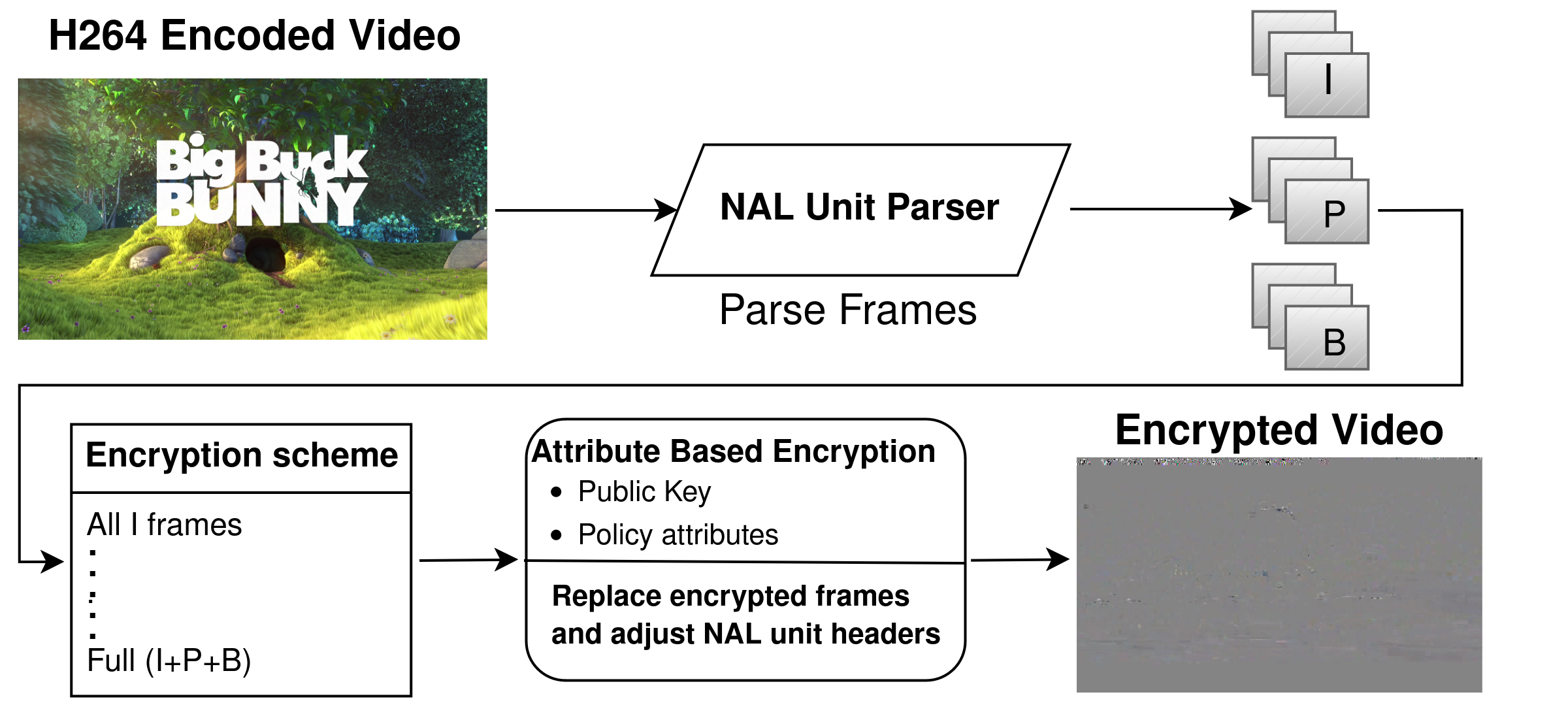}
  \captionsetup{justification=centering}
  \caption{Frame-based selective encryption using ABE.}
  \label{fig:selective-frame}
\end{figure}

Decryption mirrors this process: encrypted frame bytes are identified, decrypted using authorized keys, and reinserted into the segment. Frame size headers are restored to make the segment playable again. Figure~\ref{fig:selective-frame} illustrates our selective encryption pipeline.

% Optional claim retained for future exploration:
% Our current approach does not modify the MP4 or H264 structure, enabling format-preserving encryption~\cite{6738944}. This potentially allows in-place frame decryption and faster playback in streaming scenarios, which we aim to explore further.

\begin{figure}[h]\centering
  \includegraphics[width=.8\linewidth]{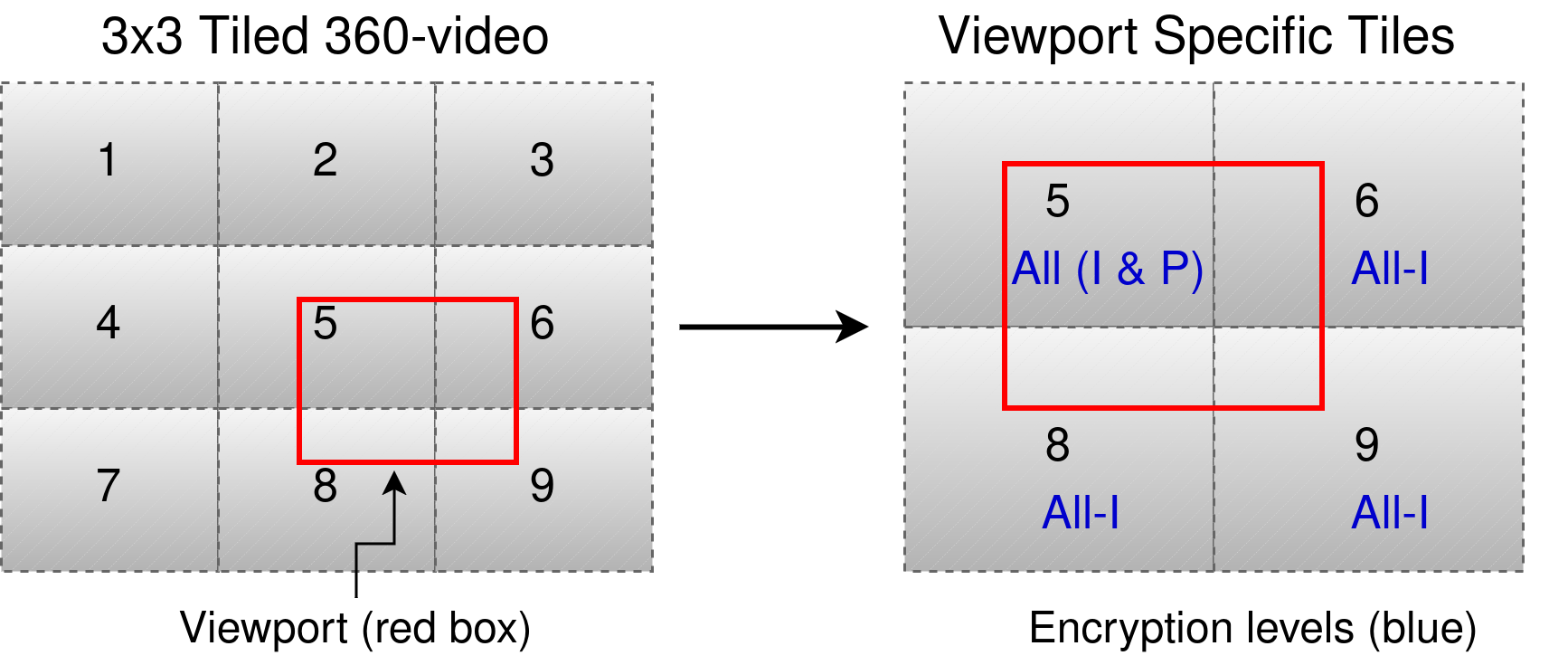}
  \captionsetup{justification=centering}
  \caption{Tile-based selective frame encryption.}
  \label{fig:tiled-enc}
\end{figure}

We extend this concept in the second dimension with \textbf{Tile-Based Selective Encryption}, which applies frame-based encryption selectively within tiles of a \tsd video based on their relevance to the user's viewport. To illustrate our approach, we use the example of a \tsd video that is tiled in a 3x3 grid. Without loss of generality, our approach can be easily extended to other tiling schemes. For tiles covering the major portion of the viewport—determined dynamically during playback—we aggressively encrypt both \textit{I }and \textit{P}-frames. For peripheral viewport tiles, we encrypt only \textit{I}-frames, achieving notable degradation in quality with reduced computational effort.

Figure~\ref{fig:tiled-enc} illustrates this process, where tile 5, occupying the core of the viewport, is fully encrypted (\textit{I }and\textit{ P-}frames), while tiles 6, 8, and 9 are partially encrypted (\textit{I}-frames only). This dual-layer strategy—spatial tiling and frame-based selection—achieves strong content protection while minimizing computational load.

% \textbf{VMAF-Based Evaluation:} To quantify the effectiveness of our encryption strategy, we use the Video Multimethod Assessment Fusion (VMAF) metric. For each experiment, we compare the encrypted version of the video against the original. For \tsd videos, we extract the user's viewport in both versions for accurate comparison. Unlike typical use cases where a high VMAF score indicates better quality, our goal is to achieve a low VMAF score, signaling effective quality degradation through selective encryption and limited computational cost.

\subsection{Prototype Implementation}
\label{subsec:protoype}
The evaluation of ABE for \tsd video streaming is based on a prototype we implemented, making use of the following components:

\textbf{CPABE Toolkit:} We implement Attribute-Based Encryption (ABE) using the CPABE toolkit~\cite{CPABE}, based on the work in~\cite{4223236}. 
%The setup phase generates a public key and master key. The public key is used to encrypt files under an attribute-based policy, while the master key is used to generate private keys for users based on their attribute sets. A user can decrypt the content if their attributes satisfy the policy defined during encryption.

\textbf{MPD:} We modified the Media Presentation Description (MPD) to support our tiled \tsd video streaming experiments using Attribute-Based Encryption (ABE). In our setup, each video segment consists of a group of four tiles representing the user’s viewport. This structure is maintained across each adaptation set (representing different video bitrates), where each segment entry represents a list of four tile files. To enable adaptive bitrate streaming, the bitrate of each adaptation set is specified as the sum of the bitrates of the four tiles. Furthermore, each tile’s filename includes a suffix that reflects the level of selective encryption (e.g., \textit{allI} or \textit{allI+P}) applied to it. This informs the client which frame types (\textit{I} or \textit{P}) are encrypted, guiding it to perform the corresponding ABE decryption during playback. 

% Listing~\ref{lst:mpd-pseudocode} shows part of an MDP file that highlights the changes to support selective encryption. 

% \lstset{
%   basicstyle=\ttfamily\footnotesize,
%   breaklines=true,
%   breakatwhitespace=true,
%   columns=fullflexible,
%   frame=single,
%   backgroundcolor=\color{gray!10},
%   captionpos=b
% }

% \begin{lstlisting}[language=XML, caption={Pseudocode of Modified MPD Structure}, label={lst:mpd-pseudocode}]
% <MPD mediaPresentationDuration="...">
%   <AdaptationSet contentType="video" id="X" bandwidth="SUM_4_TILES">
%     <Representation id="0" mimeType="video" codecs="avc">
%       <SegmentList timescale="..." duration="...">
%         <Initialization sourceURL="init-streamX.m4s" />
%           <SegmentURL media="tile1-enc_allP_chunk-streamX-00001.m4s, tile2-enc_allI_chunk-streamX-00001.m4s, tile3-enc_allI_chunk-streamX-00001.m4s, tile4-enc_allI_chunk-streamX-00001.m4s" />
%           ...
%       </SegmentList>
%     </Representation>
%   </AdaptationSet>
% </MPD>
% \end{lstlisting}

\textbf{AStream-360:} We extend the AStream DASH client, a Python-based DASH emulation tool~\cite{10147938, Juluri2015SARA} supporting various ABR algorithms, to support \tsd tiled video streaming with selective encryption. 
% Our work builds on prior modifications introduced in~\cite{10.1145/3712676.3714450}, where AStream was adapted to support HTTP-ABE by integrating attribute-based decryption.
For each playback segment, AStream-360 downloads a set of four tiles corresponding to the user’s current viewport and invokes the ABE selective decryption module to decrypt targeted frames within each tile. All four tiles are fetched at the same quality level, selected by the basic ABR algorithm based on the combined download and decryption time across all tiles. For simplicity, only viewport-relevant tiles are downloaded in our current implementation; however, this design can be easily extended to include non-viewport tiles to support seamless viewport transitions.

\section{Evaluation} \label{sec:evaluation-1}
In this section, we first analyze the impact of our two-dimensional selective encryption technique under ABE for tiled \tsd videos.
We then describe the experimental setup and the metrics used to assess the performance of our attribute-based selective end-to-end encryption framework. Finally, we present experimental results comparing our approach to HTTPS-based \tsd video streaming.

\subsection{Impact of Selective Encryption under ABE}
\label{subsubsec:enc_impact}
In this section, we apply frame-based selective encryption to tiled \tsd videos using ABE under four different strategies. We evaluate each strategy in terms of the level of content unviewability it achieves and the associated computational trade-offs. For simplicity, our ABE policy used a single attribute.

\textbf{Encryption Scheme:} We evaluate four tile-based selective encryption strategies for \tsd video streaming. Two of these employ viewport-aware, 2-dimensional selective encryption, where we encrypt frames in a tile segment based on the portion of viewport it covers: \textbf{(a) Major-allP:} All \textit{I}- and \textit{P}-frames are encrypted for the tile that predominantly covers the viewport (e.g., tile 5 in Fig.~\ref{fig:tiled-enc}), while only \textit{I}-frames are encrypted for the remaining minor tiles in the viewport. \textbf{(b) Major-allI:}  Only \textit{I}-frames from the major tile are encrypted, and all minor tiles in a viewport remain unencrypted.

To generate encrypted viewport videos, we identify the four tiles present in the viewport for each segment. Based on whether a tile is classified as major or minor, we apply the corresponding encryption level. These tiles are combined and cropped to form a viewport-specific clip for each segment. This process is repeated for all segments, and the clips are concatenated to produce the final encrypted viewport video.

The remaining strategies apply encryption independent of the viewport: \textbf{(c) Full:} All frames (\textit{I, P, and B}) are encrypted in each tile. \textbf{(d) All I+P:} Only all \textit{I}- and all \textit{P}-frames are encrypted in each tile.

\textbf{360-Video Dataset:} Our evaluation uses four diverse \tsd videos: (1) an action-packed short film, (2) a London tour with relatively still backgrounds, (3) a slower-paced documentary with frequent motion and dynamic background changes, and (4) a yoga session with minimal motion and a static indoor setting. These videos were selected to represent various genres and motion complexities.

\textbf{Tiling and Encoding:} Each video is divided into a 3x3 grid, resulting in nine tiles. With a resolution of 3840x1920, each tile measures 1280x640 pixels. Tiles are encoded using MP4-H264 and segmented for DASH streaming. A keyframe interval of 60 frames is used to support ABR, with additional keyframes inserted at scene changes using a scene threshold of 40.

\textbf{Viewport Simulation:} We assume the viewport corresponds to the size of a single tile, approximately 120 degrees of the horizontal field of view (360/3). Since real head movement data was unavailable for most videos, we manually analyzed each video to identify visually engaging areas likely to attract user attention. A viewport spans portions of four different tiles and changes to different tiles at various points in the video. The viewport portions are cropped and stitched into a continuous "viewport video" for further analysis.

\textbf{VMAF Evaluation:} To assess quality degradation due to selective encryption, we compare each encrypted viewport video to its unencrypted counterpart using VMAF. In our context, a lower VMAF score indicates effective quality degradation, aligning with the goal of restricting high-quality video access while minimizing computational overhead.

\begin{figure*}[t]
    \centering
    \begin{subfigure}[b]{0.3\textwidth}
        \centering
        \includegraphics[width=\textwidth]{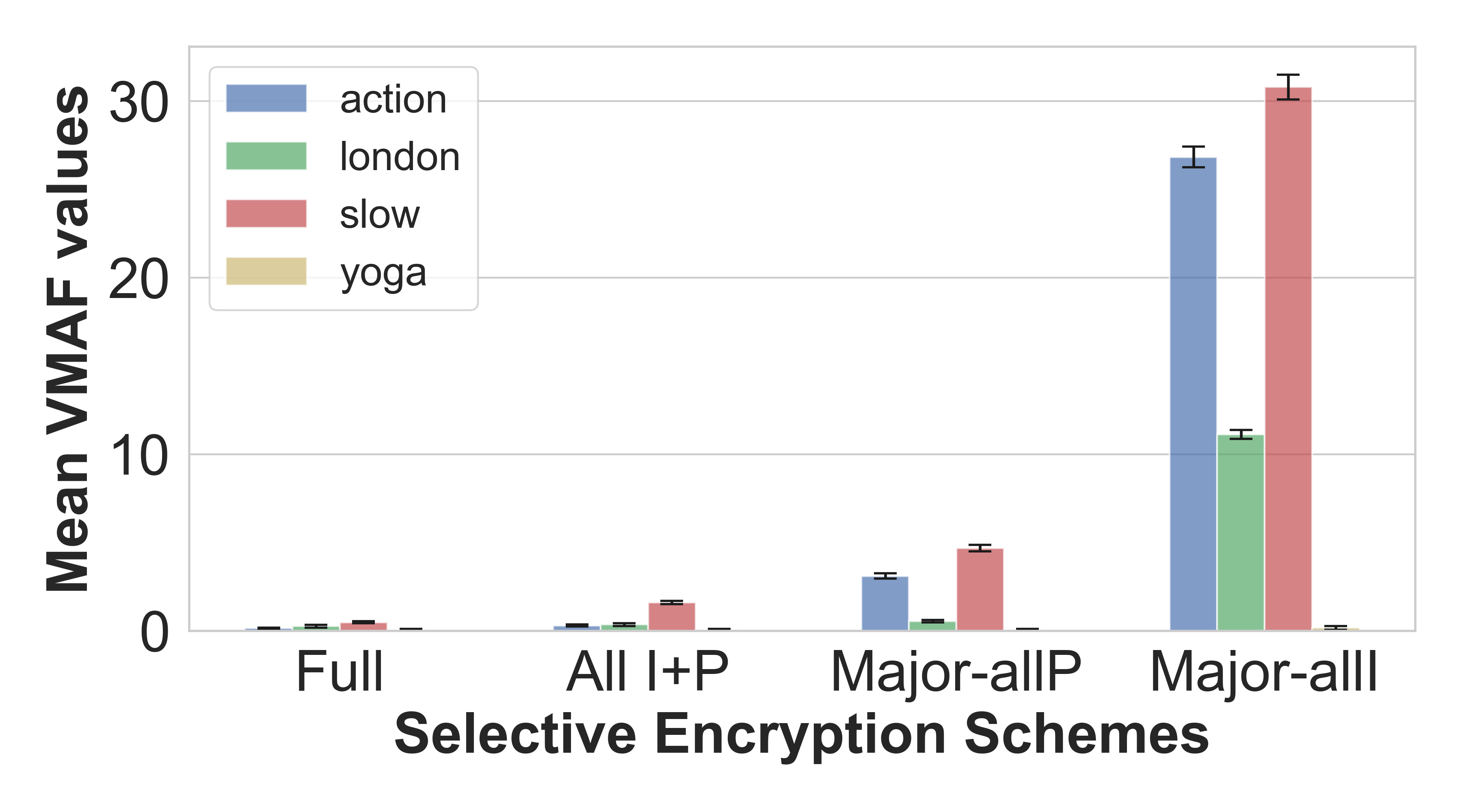}
        \caption{Average VMAF}\label{fig:vmaf-time-overhead:a}
    \end{subfigure}
    \hfill
    \begin{subfigure}[b]{0.3\textwidth}
        \centering
        \includegraphics[width=\textwidth]{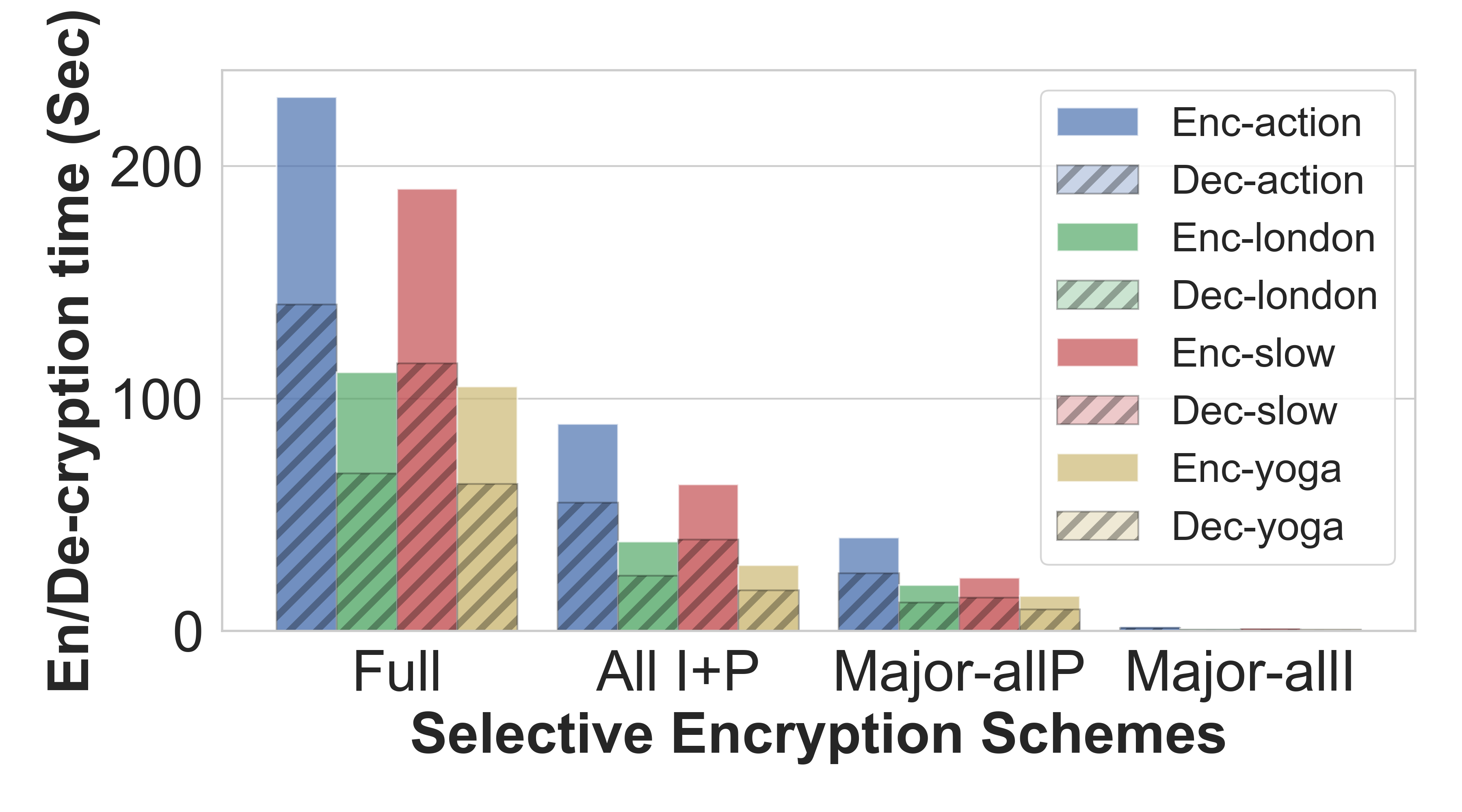}
        \caption{ Encryption \& Decryption times}\label{fig:vmaf-time-overhead:b}
    \end{subfigure}
    \hfill
    \begin{subfigure}[b]{0.3\textwidth}
        \centering
        \includegraphics[width=\textwidth]{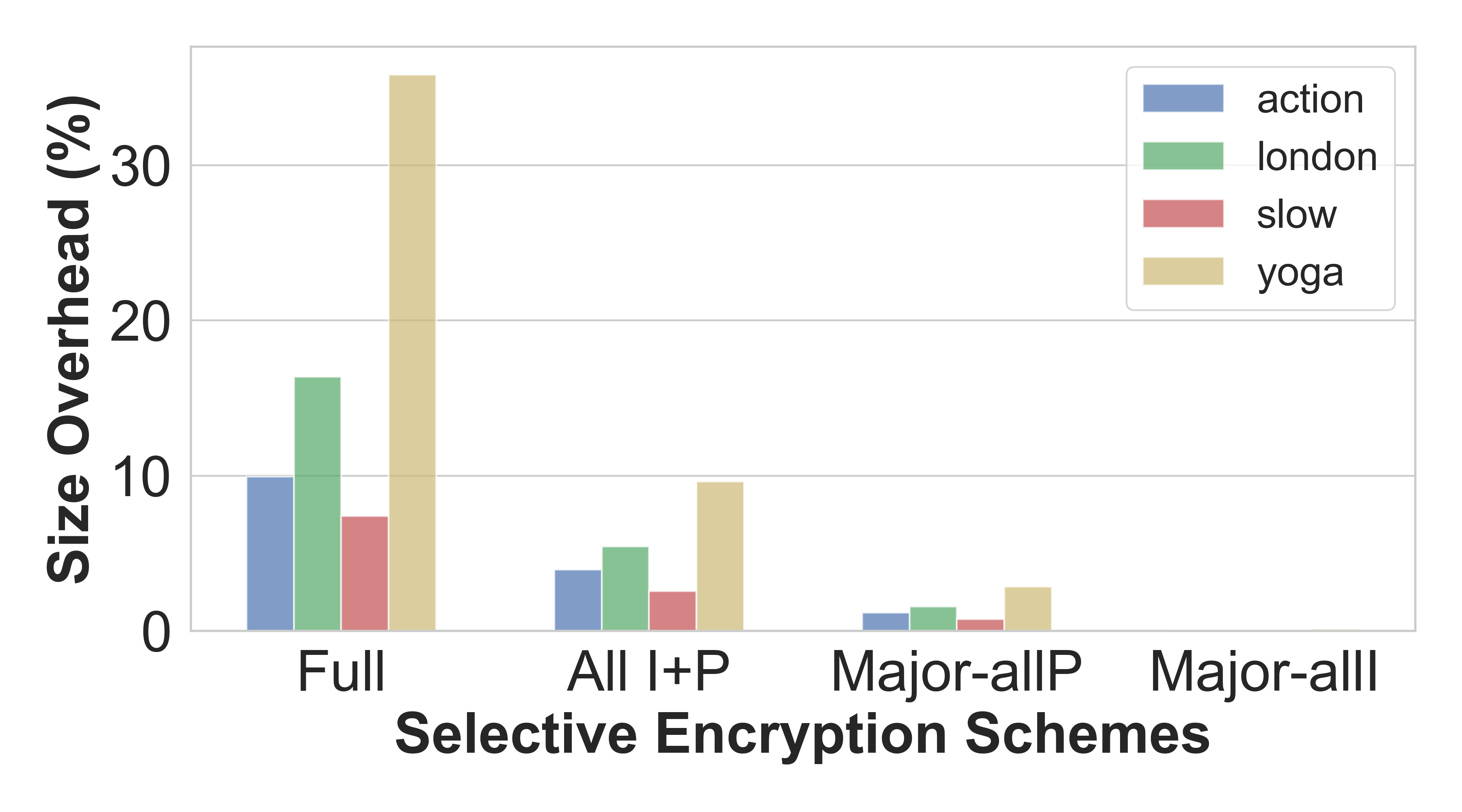}
        \caption{Percent of size overhead}\label{fig:vmaf-time-overhead:c}
    \end{subfigure}
 %   \vspace{-.4cm}
    \caption{Impact of different selective encryption schemes under ABE.}
    \label{fig:vmaf-time-overhead}
%    \vspace{-.4cm}
\end{figure*}

Figure~\ref{fig:vmaf-time-overhead:a} presents the mean VMAF scores with 95\% confidence intervals for the four selective encryption schemes. As expected, Full encryption results in the lowest VMAF, making content unviewable. Both All I+P and Major-P also achieve strong degradation, keeping VMAF consistently below 5. In contrast, Major-allI performs the weakest, with VMAF reaching up to 30, as it encrypts only the \textit{I}-frames of the major tile but leaves the rest unencrypted.

An interesting exception is the Yoga video, where VMAF remains near zero across all schemes, likely due to its static content and minimal motion. These findings suggest that encryption strategies can be content-aware, adapting frame selection based on video complexity to balance degradation and computational overhead.

% \begin{figure}[h]\centering
%   \includegraphics[width=.9\linewidth]{figs/vmaf-viewport.png}
%   \captionsetup{justification=centering}
%   \caption{Average VMAF for different Tile-Based Selective Encryption schemes for 360-videos}
%   \label{fig:vmaf-viewport}
% \end{figure}

\textbf{Encryption \& Decryption time overhead:} Figure~\ref{fig:vmaf-time-overhead:b} shows the total encryption and decryption time for each scheme. Full encryption incurs the highest runtime, followed by All I+P, which reduces the cost to half by excluding \textit{B}-frames. Major-allP cuts runtime even further—about half of All I+P—while still keeping VMAF below 5. Major-allI is the most efficient, with runtimes of only a few seconds, but it delivers the weakest degradation.

% \textbf{Comparisons with AES?:

\textbf{Total size increase overhead:} Figure~\ref{fig:vmaf-time-overhead:c}, reports the size overhead introduced by each scheme. As with runtime, the Full scheme results in the largest overhead, most notably over 35\% for the yoga video. All I+P keeps overhead moderate, under 10\% only, while Major-allP limits it to around 5\% by restricting encryption to select tiles and frames. Major-allI introduces negligible overhead, typically under 1\%, aligning with its minimal computational cost.

% \textbf{average size increase per frame:}
% \textbf{Encryption \& Decryption Times and VMAF Analysis of Individual Tiles:} this can potentially add 2 set of 9 fig for each tiles not sure about this since it takes too much space on paper.

\subsection{\tsd Video Streaming Setup \& Metrics}
\label{subsec:setupandm}
\subsubsection{Setup}
\label{subsubsec:setup}

\textbf{Videos:} 
For our streaming experiments, we used Google Spotlight Stories: HELP, the same action video referenced in Sect.~\ref{subsubsec:enc_impact}.  The 360-degree video, originally in 3840×1920 (4K) resolution, was divided into a 3×3 tile grid, resulting in nine tiles. Each tile was encoded using the H.264/AVC codec~\cite{ITU_H.264}, packaged in MP4, and segmented into MPEG-DASH compliant .m4s files.

To enable adaptive bitrate streaming, we generated four quality levels per tile, each with distinct resolutions and bitrates (Tab.~\ref{tab:stream_qualities}). Streams were encoded with a 60-frame keyframe interval, with additional keyframes inserted at scene changes (threshold = 40). We prepared two dataset versions with 2-second and 4-second segments for comparative evaluation. The 293-second video yields 147 segments for the 2-second version and 74 segments for the 4-second version per quality level. To create a more diverse dataset, we created four symbolic links for each tiled video stream, effectively simulating five distinct video copies.

A key reason for selecting this video was the availability of real head movement data from 48 viewers~\cite{10.1145/3083187.3083210}. We parsed each user’s head trace to extract the four tiles forming the viewport per segment, designating the tile covering the largest area as the Major tile and the rest of them as Minor tiles.

We averaged viewport coverage across users for both 2-second and 4-second segments to generate per-segment tile selections. As described in Sect.~\ref{subsec:protoype}, these were encoded into the MPD, with each segment referencing four tiles—one major and three minor—based on real viewing data. The 2-second segmentation yielded finer viewport accuracy due to less averaging loss. We created 40 unique MPD files from 40 user head traces and evenly distributed five simulated video copies (via symbolic links) across them. To reflect realistic content popularity, clients requested videos according to a Zipf distribution ($s=1.5$), ensuring more frequent access to certain videos—mirroring real-world streaming patterns.
% and increases the likelihood of popular videos being cached.

\begin{table}[h!]
\small
\centering
\caption{Bitrates and resolutions per video tile.}
\begin{tabular}{|c|c|c|}
\hline

\textbf{Stream} & \textbf{Bitrate (Mbps)} & \textbf{Resolution} \\ \hline
1               & 0.5                     & 480x240             \\ \hline
2               & 1                    & 640x320             \\ \hline
3               & 2                    & 960x480             \\ \hline
4               & 3                    & 1280x640            \\ \hline
\end{tabular}
\label{tab:stream_qualities}
% \vspace{-.65cm}
\end{table}

\textbf{ABE:} 
As part of our evaluation, we applied ABE encryption to all tiled video segments across all quality levels, using a single-attribute policy for simplicity. While prior work~\cite{reddick2022wip} shows that decryption time grows linearly with the number of attributes, its overall impact on system performance is minimal. Following encryption, we updated stream bitrates to account for overhead and revised the MPD files with the new bitrates and encrypted segment URLs. We evaluated two selective encryption schemes: (1) Major-P, which applies viewport-aware, two-dimensional encryption, and (2) All-I+P, which encrypts frames regardless of viewport coverage (see Sect.~\ref{subsubsec:enc_impact} for details).
    
We conducted our experiments on the \textbf{CloudLab} testbed~\cite{duplyakin2019cloudlab} using two deployment scales.

% following the methodology from~\cite{10.1145/3712676.3714450}.

\textbf{(a) Small-Scale Setup:}
This setup consists of seven nodes: one origin server, one cache, and five client nodes. Each client node runs six DASH clients, totaling 30 clients, with each client using a unique headtrace-based MPD file. Bandwidth limits were configured using Linux’s tc utility~\cite{tc}. With each client streaming four tiles at up to 3 Mbps per tile (12 Mbps total), each node requires up to 72 Mbps, resulting in 360 Mbps across all clients. To avoid client-side contention, we allocated 72 Mbps between the cache and each client node. To stress the origin-cache link, we limited its bandwidth to 120 Mbps—one-third of the total client bandwidth. 

\textbf{(b) Large-Scale Hierarchical Setup:}
This setup spans eight nodes arranged in a two-level caching hierarchy. One node hosts the origin server, another acts as the level-1 (L1) cache, and two level-2 (L2) caches connect to the L1 cache. Each L2 cache connects to two client nodes, with each client node running 20 clients, totaling 80 clients. Since we had 40 unique headtraces, each was reused once. Bandwidth limits were enforced similar to the small-scale setup, between each L2 cache and its client nodes was capped at 240 Mbps (480 Mbps total). To simulate contention, we limited the L1-to-L2 links to 160 Mbps each (one-third of downstream capacity). The link between the origin and L1 cache was set to 320 Mbps, matching the combined bandwidth from L1 to both L2 caches.

\textbf{Apache HTTP Server \& Apache Traffic Server:}  We use the Apache HTTP Server~\cite{apache2} on the origin server and Apache Traffic Server (ATS)~\cite{ATS} as the caching proxy. Both were configured with TLS/SSL to support HTTPS streaming. In the small-scale experiment, cache sizes were varied across seven configurations: 0 MB (disabled), 100 MB, 250 MB, 500 MB, 1000 MB, 1500 MB, and 2000 MB. In the large-scale hierarchical setup, a minimum cache size of 10 MB was used, as 0 MB caching is not supported by ATS.

\textbf{DASH Clients:} For client-side streaming, we use AStream-360 (see Sect.\ref{subsec:protoype}), which includes ABE decryption for HTTP-ABE experiments. For HTTPS, the same client is used without decryption logic. Client session start times are scheduled using a Poisson distribution\cite{poisson}. In the small-scale setup, we set $\lambda = 20$, resulting in an average interval of 20 seconds between client starts. For the large-scale setup, we increase the rate to $\lambda = 10$ to reduce total experiment time, yielding an average interval of 10 seconds.

\subsubsection{Evaluation Metrics,}

\textbf{CPU Load:}
To compare the computational overhead of HTTP-ABE and HTTPS, we record CPU usage on both the cache and origin servers. We use Linux’s pidstat tool~\cite{pidstat} with a one-second refresh interval to monitor CPU utilization in real time throughout each experiment.

\textbf{Hit Rate:}
We assess cache efficiency by comparing hit rates under HTTP-ABE and HTTPS across various cache sizes. Hit rates are computed using ATS’s monitoring logs, which report the number of cache hits and misses during streaming sessions.

% \textbf{Midgress Traffic:}
% Midgress refers to traffic between a cache and the origin server due to cache misses. We evaluate midgress traffic under both HTTP-ABE and HTTPS by varying cache sizes and completing full streaming sessions for all clients. To measure this traffic, we use tcpdump~\cite{tcpdump} on the cache, filtering packets to and from the origin server. We then analyze the captured data with capinfos~\cite{capinfos} to determine the total data volume. In the large-scale hierarchical experiment, we additionally measure midgress between each L2 cache and the L1 cache.

\textbf{QoE-1: VMAF (Video Quality):}
To evaluate perceived video quality, we use VMAF~\cite{vmaf}, a metric developed by Netflix that combines human visual models with machine learning. VMAF scores range from 0 to 100, with higher scores indicating better visual quality. For each client, we compare the streamed quality of a tile segment to the corresponding tile segment from the highest-quality stream. We compute the average VMAF per client and then aggregate scores across all clients to assess overall video quality performance across cache sizes.

\textbf{QoE-2: Rebuffering:}
Rebuffering measures playback interruptions caused by an empty buffer. We extract rebuffering durations from the AStream DASH client logs, which record segment-wise stall times. These metrics are aggregated per client and subsequently averaged across all clients to assess the impact of caching and protocol selection (HTTP-ABE vs. HTTPS) on playback smoothness.

\begin{figure}[t]
    \centering
    \begin{subfigure}[b]{\linewidth}
        \centering
        \includegraphics[width=0.75\linewidth]{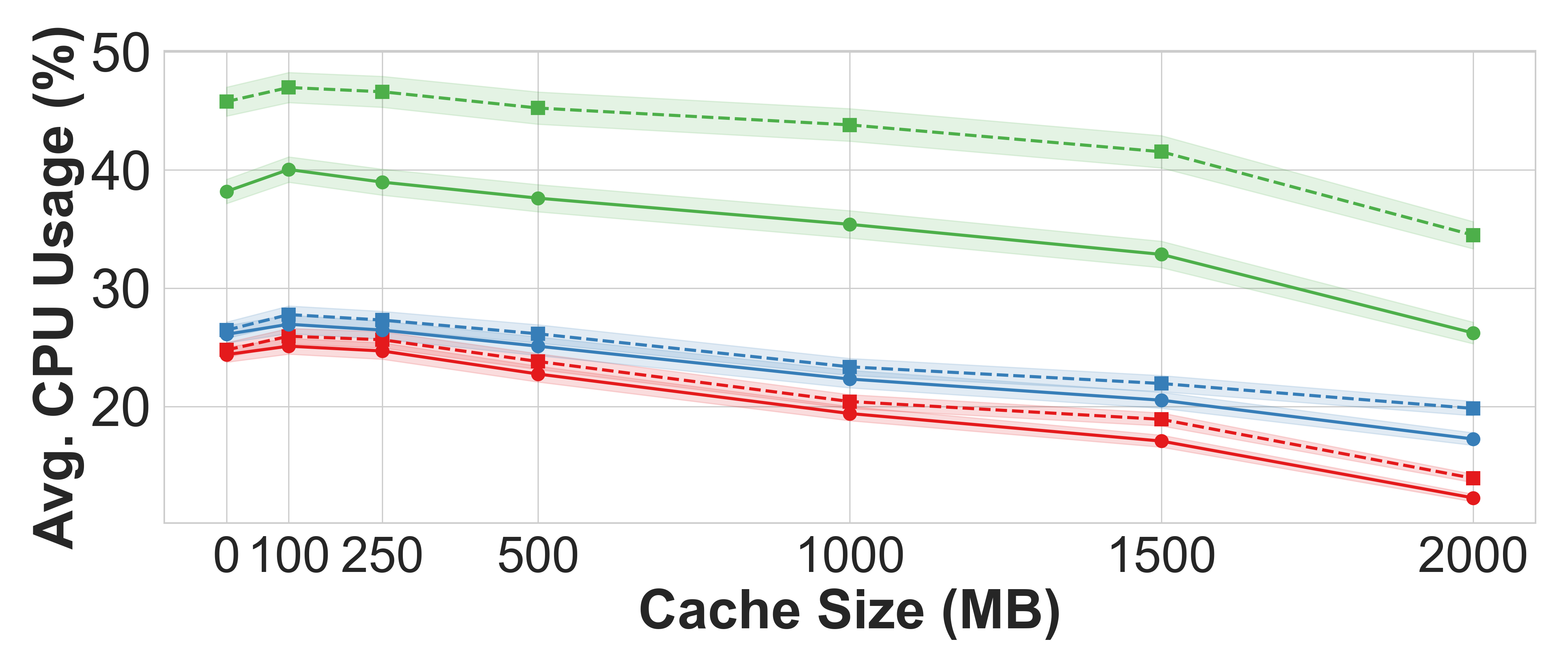}
        \caption{Cache CPU usage (keys are similar to Fig.~\ref{fig:smallscale-hitrate})}
        \label{fig:smallscale-cpu-cache}
    \end{subfigure}

    \vspace{1em} % optional spacing between plots
    
    \begin{subfigure}[b]{\linewidth}
        \centering
        \includegraphics[width=0.75\linewidth]{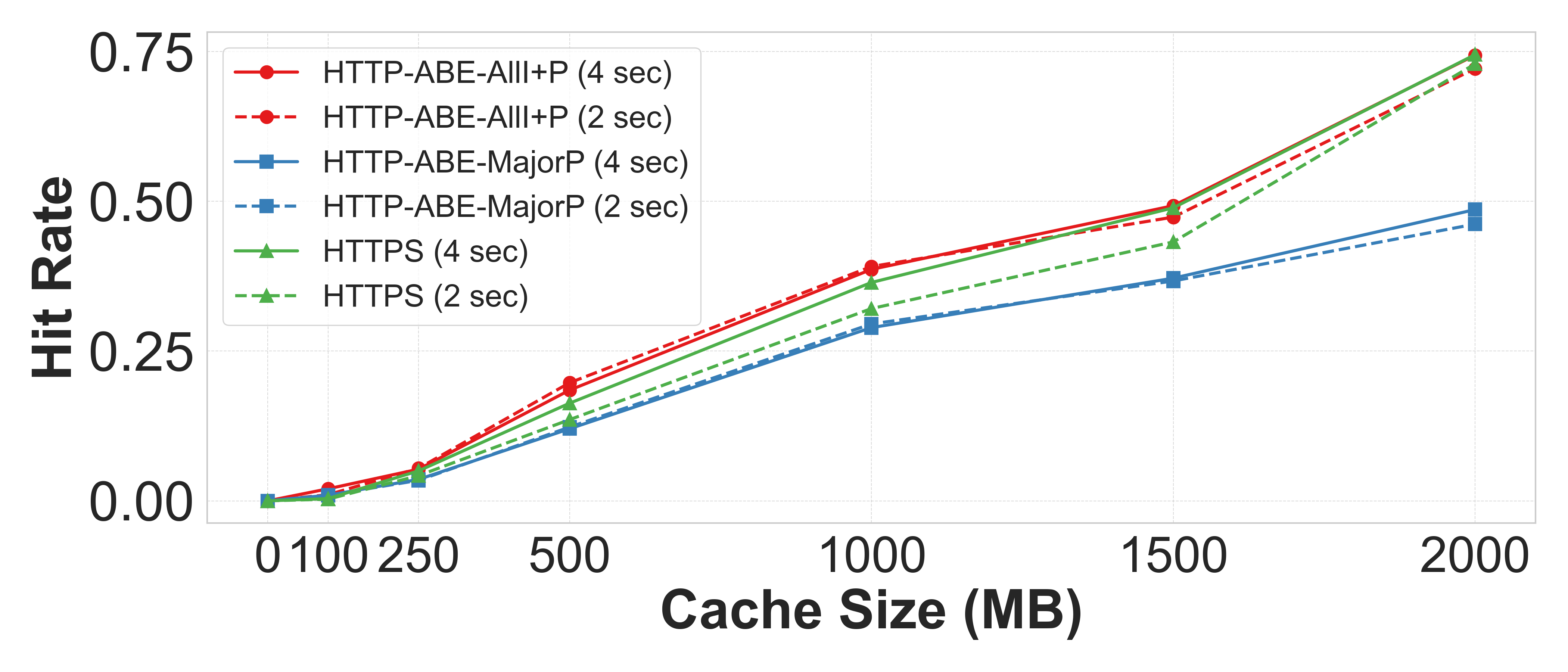}
        \caption{Cache hit rate}
        \label{fig:smallscale-hitrate}
    \end{subfigure}
  %  \vspace{-.5cm}
    \caption{Comparison of cache performance metrics for the small-scale experiment.}
    \label{fig:cache-hitrate}
\end{figure}

\subsection{HTTP-ABE vs HTTPS \tsd video streaming} \label{abe-vs-https}
To evaluate system performance, we analyzed cache and origin server CPU load, cache hit rate, midgress traffic, and client QoE. Two experiments were conducted: a small-scale setup and a large-scale hierarchical multi-cache setup, as described in Sect.~\ref{subsubsec:setup}.

The small-scale setup included thirteen runs: one with 0 MB cache and two each for 100 MB, 250 MB, 500 MB, 1000 MB, 1500 MB, and 2000 MB. In the large-scale setup, where 0 MB was unsupported, we used 10 MB as the minimum, resulting in fourteen runs. Cache sizes were adjusted simultaneously across all three caches (L1 and both L2s). Each experiment included a warm-up run to populate the cache (no metrics collected), followed immediately by a second run that records metrics for comparison. 

% The third experiment uses the same large-scale hierarchical setup but removes bandwidth limits to eliminate traffic congestion. This adjustment allows us to focus solely on comparing the performance of the two approaches—HTTP-ABE and HTTPS under unconstrained network conditions.

% %\vspace{-.12in}

\subsubsection{\textbf{Small-Scale Experiment}}
\label{subsubsec:smallscaleexp}
In this experiment, we evaluate the performance of our two ABE-based selective encryption schemes, MajorP and All I+P, under HTTP and compare them with HTTPS-based streaming. The evaluation is conducted using two segment durations: 2-second and 4-second video segments, to study the impact of segment granularity on system performance.

\textbf{CPU Load on Cache:} In Fig.~\ref{fig:smallscale-cpu-cache}, we plot the average CPU usage throughout the streaming session, along with a 95\% confidence interval, for various cache sizes. The results show that both HTTP-ABE approaches consume significantly less CPU resources than HTTPS, with HTTP-ABE-allI+P always outperforming Major-P.

A notable performance gap appears when comparing 4-second and 2-second segment lengths. With HTTPS, the CPU load increases substantially for 2-second segments, as the cache must perform TLS termination, decryption, and re-encryption for twice the number of segments compared to the 4-second setup. In contrast, both ABE-based approaches exhibit relatively stable CPU usage, with only a slight increase when moving to shorter segments—thanks to the absence of TLS termination. \emph{This trend suggests that as segment duration decreases, HTTP-ABE becomes increasingly more efficient than HTTPS in terms of CPU usage}.

Specifically, for 4-second segments, HTTP-ABE-allI+P uses 36\% less CPU at 0MB cache and up to 53\% less at 2000MB, compared to HTTPS. For 2-second segments, the savings increase, ranging from 45\% (at 0MB) up to 60\% (at 2000MB). The HTTP-Major-P scheme shows similar trends, using 31\% up to 34\% less CPU than HTTPS for 4-second segments, and 42\% up to 47\% less for 2-second segments, as cache size increases from 0MB up to 2000MB.

% \begin{figure*}[t]
%     \centering
%     \begin{subfigure}[b]{0.45\textwidth}
%         \centering
%         \includegraphics[width=0.7\textwidth]{figs/smallscale-cache_cpu_usage_comparison.png}
%         \caption{Cache CPU usage (keys are similar to Fig.~\ref{fig:smallscale-hitrate})}\label{fig:smallscale-cpu-cache}
%     \end{subfigure}
%     \hfill
%     \begin{subfigure}[b]{0.45\textwidth}
%         \centering
%         \includegraphics[width=0.7\textwidth]{figs/smallscale-hitrate_plot.png}
%         \caption{Cache hit rate}\label{fig:smallscale-hitrate}
%     \end{subfigure}
    
%     \caption{Comparison of cache performance metrics for small-scale experiment.}
%     \label{fig:cache-hitrate}
% \end{figure*}

\begin{figure*}[t]
    \centering
    \begin{subfigure}[b]{0.32\textwidth}
        \centering
        \includegraphics[width=0.95\textwidth]{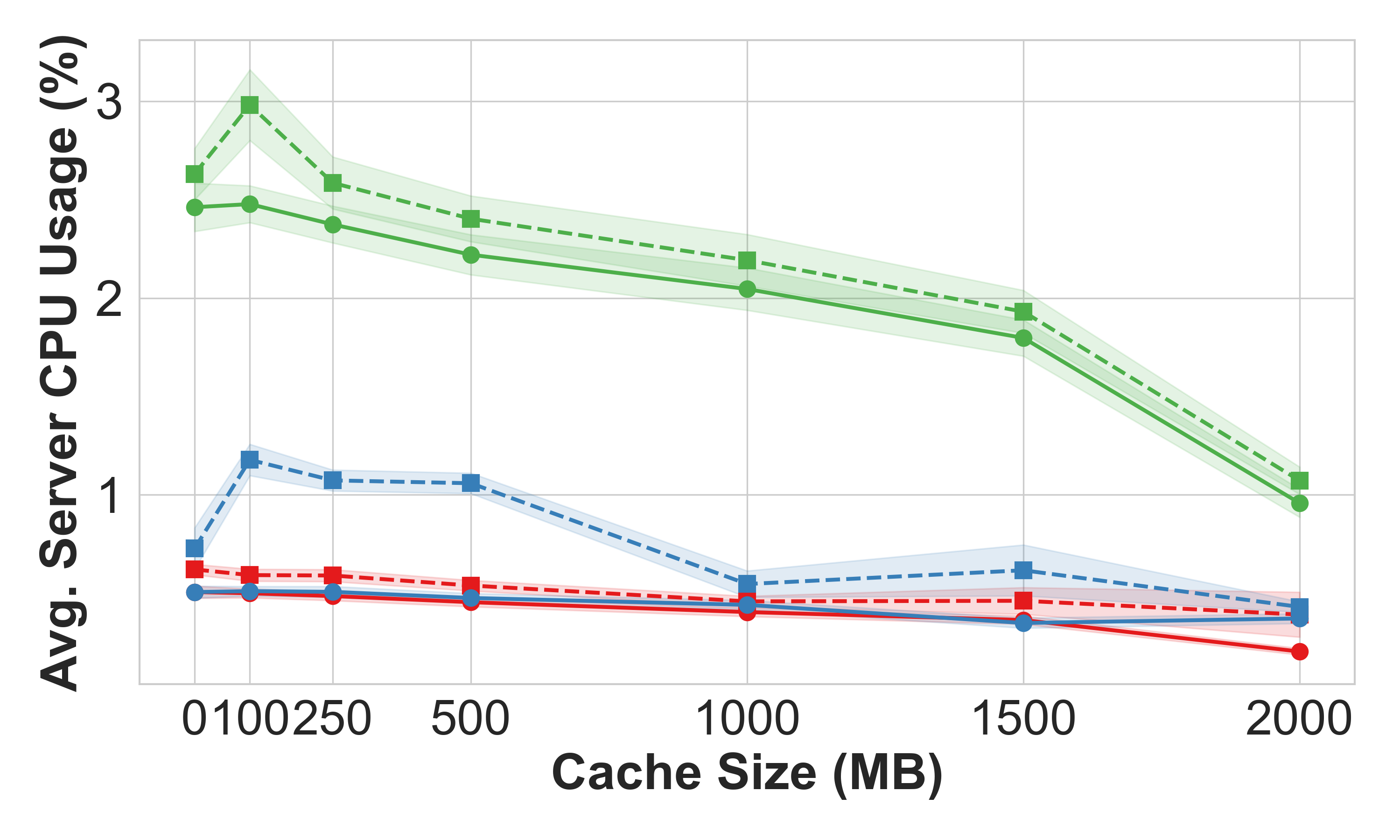}
        \caption{Server CPU usage (keys similar to Fig.~\ref{fig:smallscale-rebuffer})}\label{fig:smallscale-cpu-server}
    \end{subfigure}
    \hfill
    \begin{subfigure}[b]{0.32\textwidth}
        \centering
        \includegraphics[width=0.95\textwidth]{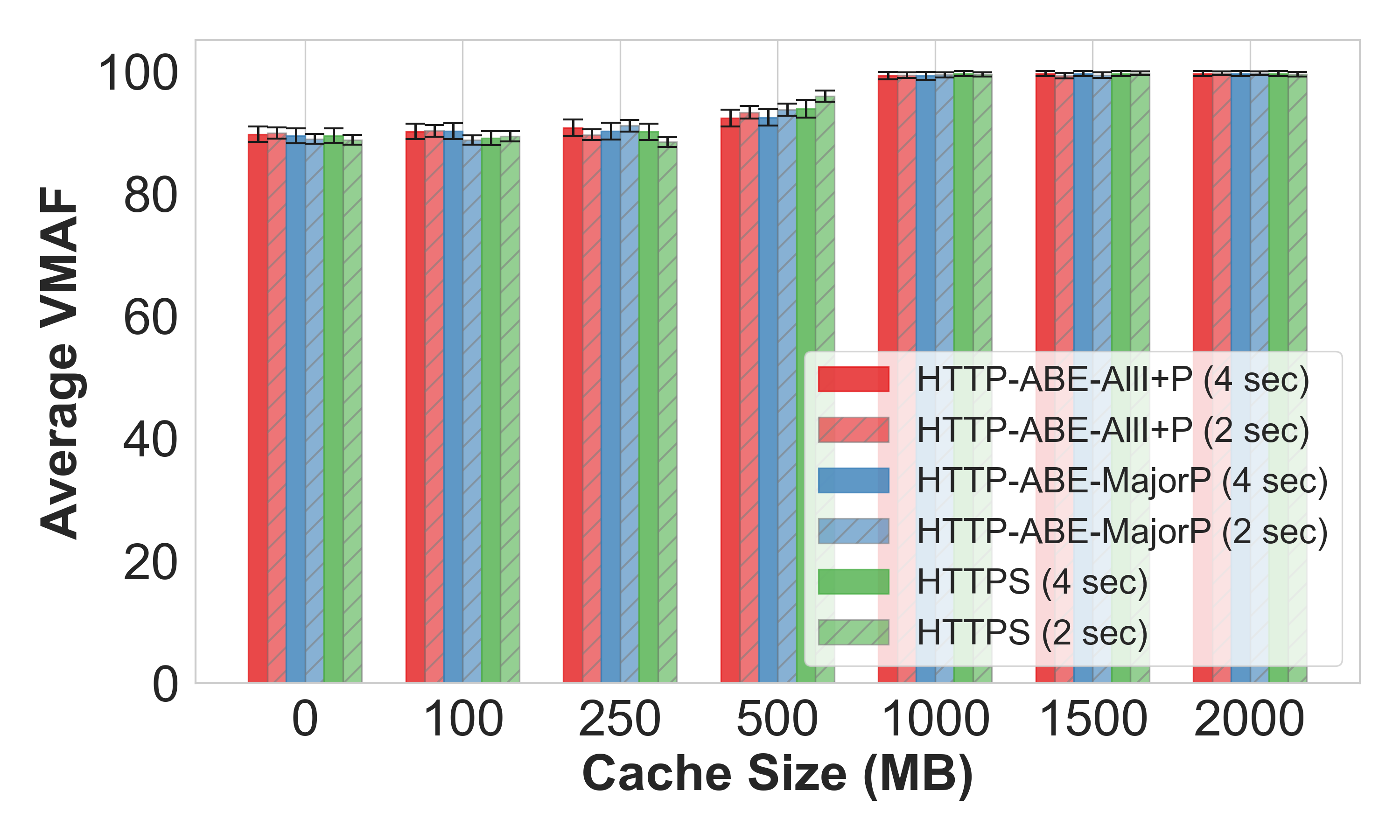}
        \caption{Average VMAF}\label{fig:smallscale-VMAF}
    \end{subfigure}
    \hfill
    \begin{subfigure}[b]{0.32\textwidth}
        \centering
        \includegraphics[width=0.95\textwidth]{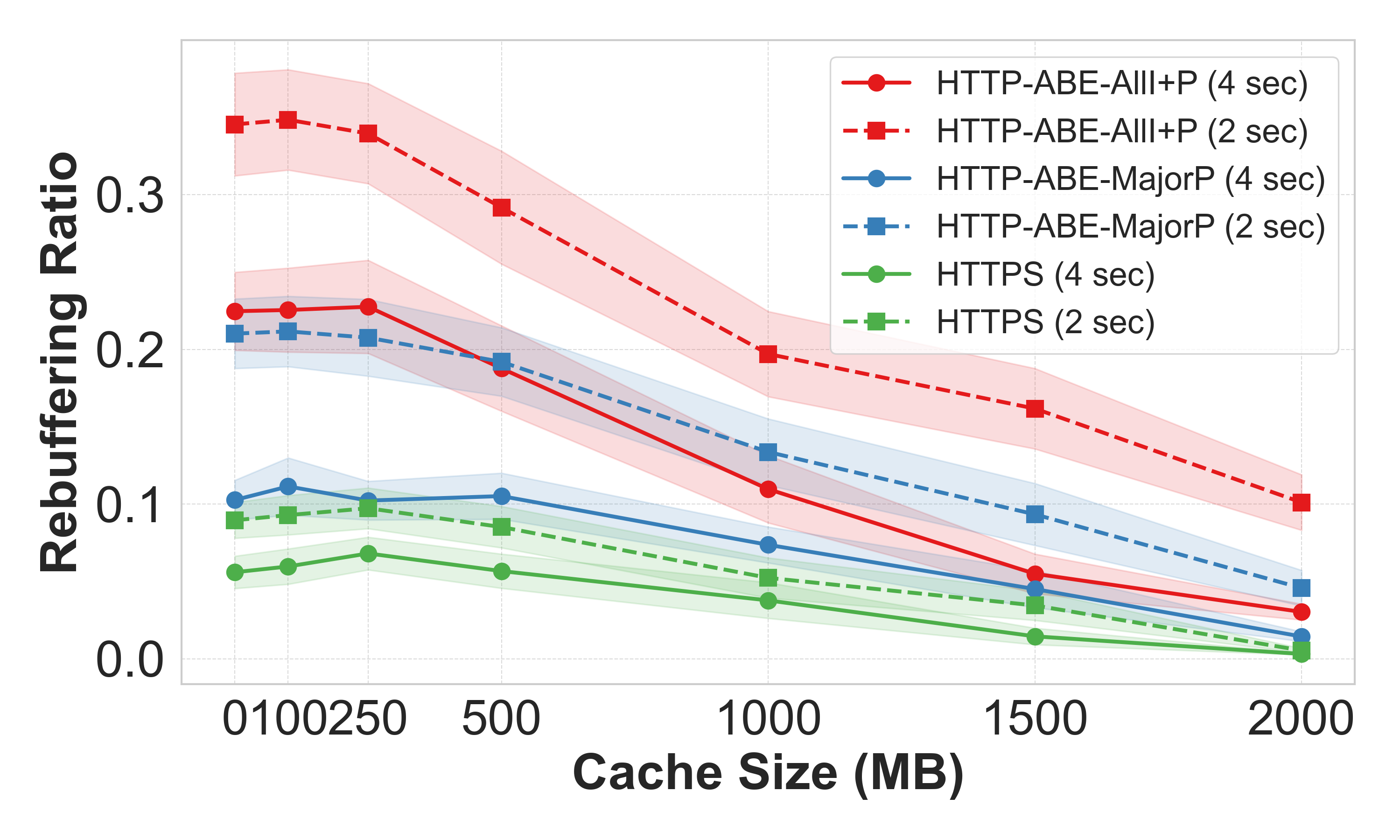}
        \caption{Rebuffering ratio}\label{fig:smallscale-rebuffer}
    \end{subfigure}
    
    \caption{Comparison of origin server CPU usage and client side QoE metrics for small-scale experiment.}
    \label{fig:server-VMAF-rebuffer}
\end{figure*}

\textbf{Comparison of Hit Rate:} Figure~\ref{fig:smallscale-hitrate} compares the hit rates of the cache for different sizes for HTTP-ABE and HTTPS. An interesting observation is that the ABE-allI+P scheme achieves higher hit rates than ABE-MajorP for cache sizes 500MB onwards. This is attributed to the nature of two-dimensional selective encryption in MajorP: the same tile segment may be stored in the cache encrypted with only I-frames (as a minor tile in one request), but later requested with both I and P-frames encrypted (as a major tile in another request), resulting in a cache miss due to mismatch in encryption level. This fragmentation leads to lower cache reuse. In contrast, the ABE-allI+P scheme applies a uniform encryption level across all tiles, improving cacheability and hit rate consistency.

This also explains the higher CPU usage for ABE-MajorP compared to ABE-allI+P, as more cache misses require more frequent segment retrieval from the origin and additional writes to the cache.

When comparing segment lengths, both ABE schemes show stable hit rates between 2-second and 4-second segments. However, HTTPS experiences a noticeable drop in hit rates for 2-second segments, particularly at cache sizes between 500 MB and 1500 MB, where ABE-allI+P outperforms HTTPS. \emph{Cache logs reveal that the ABE approach benefits from a higher number of Read While Write (RWW) hits, meaning the cache can immediately begin serving a segment to other clients while it is still being written}. In contrast, HTTPS incurs additional delays due to TLS termination, decryption, and re-encryption—even for cached segments.

This behavior becomes more pronounced with shorter segments, where client requests are more frequent, increasing the overhead of TLS operations in HTTPS. In contrast, HTTP-ABE avoids this overhead entirely, resulting in better hit rates and improved efficiency, especially with 2-second segmentation.

% \textbf{Comparison of Midgress Traffic:} Figure~\ref{fig:cache-hitrate-midgress} (right) shows that midgress traffic decreases as cache size increases, because more segments are retained in the cache. When comparing the three approaches, HTTP-ABE-allI+P and HTTPS exhibit similar midgress patterns for both 2-second and 4-second segment durations. In contrast, HTTP-ABE-MajorP incurs noticeably higher midgress traffic at larger cache sizes (1000 MB to 2000 MB). This behavior mirrors the hit rate results discussed earlier—due to variable encryption levels applied to the same tile segment, cache misses occur more frequently under MajorP, leading to increased midgress traffic and higher CPU usage.

\textbf{Origin Server:} Figure~\ref{fig:smallscale-cpu-server} shows the origin server’s CPU load with a 95\% confidence interval. Across all cache sizes and for both 2-second and 4-second segments, both HTTP-ABE approaches consistently use less CPU than HTTPS. This is primarily due to the overhead of TLS encryption required for each client request under HTTPS. As cache size increases, the CPU usage gap between HTTPS and HTTP-ABE narrows, since more segments are served from the cache. However, overall CPU usage remains low for all configurations, with HTTPS peaking at approximately 3\%. 
    
\textbf{Comparison of VMAF:} We compared the client’s QoE between HTTP-ABE and HTTPS by calculating the VMAF score for each tile segment and averaging the results per client. Figure~\ref{fig:smallscale-VMAF}  presents the average VMAF across 30 clients, along with a 95\% confidence interval, for various cache sizes. The results show that all three approaches deliver comparable VMAF scores, regardless of cache size or segment duration (2-second or 4-second). Additionally, we observe that average VMAF scores are lower at smaller cache sizes and increase with larger caches, as more segments are served directly from the cache. This reduces requests to the origin server and avoids potential congestion.

\textbf{Comparison of Rebuffering ratio:} Figure~\ref{fig:smallscale-rebuffer} shows the average rebuffering ratio over the duration of the entire video for 30 clients, with a confidence interval of 95\% between varying cache sizes. We observe that HTTP-ABE-allI+P experiences significantly higher rebuffering compared to ABE-MajorP and HTTPS, particularly when using 2-second segments. The HTTPS approach consistently results in the lowest rebuffering, while MajorP performs similarly to HTTPS with 4-second segments but degrades noticeably with 2-second segments.

Rebuffering decreases for all approaches as cache size increases, since more segments are served directly from the cache. The higher rebuffering observed in HTTP-ABE approaches can be attributed to bandwidth congestion between the origin server and the cache, further exacerbated by the increased segment sizes introduced by ABE encryption overhead (as discussed in Sect.~\ref{subsubsec:enc_impact} and shown in Fig.~\ref{fig:vmaf-time-overhead:c}). This overhead is greater for the allI+P scheme compared to MajorP, which explains its poorer performance.

Similar findings were reported in~\cite{10.1145/3712676.3714450}, which shows that removing bandwidth constraints significantly reduces rebuffering for HTTP-ABE and HTTPS, limiting it to initial startup buffering.

\begin{figure*}[]\centering
    \includegraphics[width=.95\linewidth]{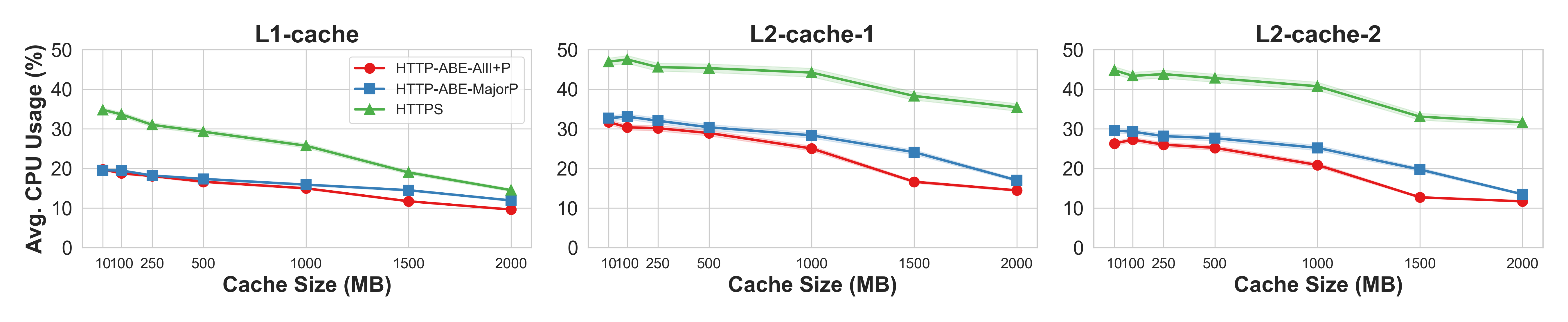}
  \captionsetup{justification=centering}
  \caption{CPU usage of L1 and L2 caches for different cache sizes.}
    \label{fig:l1l2:cpu-cache}
\end{figure*}

 \begin{figure*}[]\centering
  \includegraphics[width=.95\linewidth]{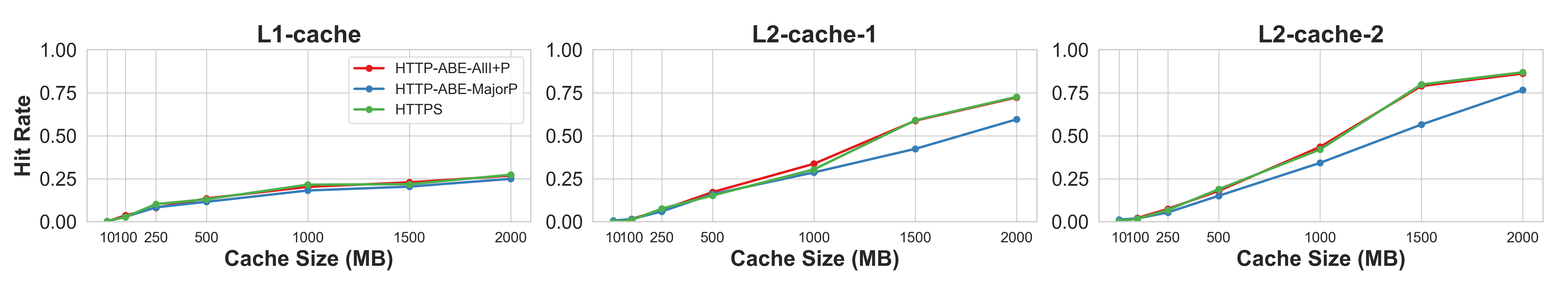}
  \captionsetup{justification=centering}
  \caption{Hitrate of L1 and L2 caches for different cache sizes.}
    \label{fig:l1l2:hitrate}
\end{figure*}

\begin{figure*}[t]
    \centering
    \begin{subfigure}[b]{0.32\textwidth}
        \centering
        \includegraphics[width=0.95\textwidth]{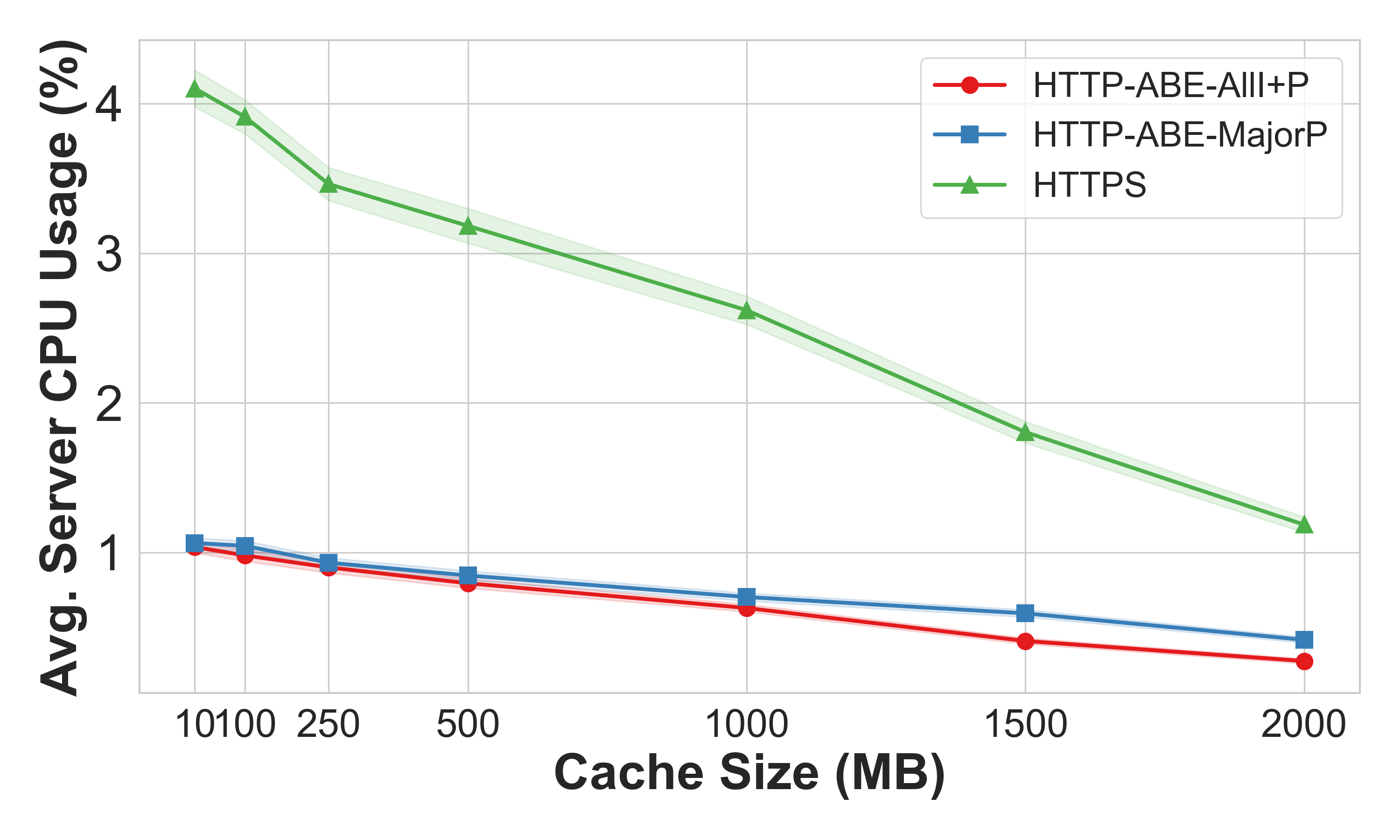}
        \caption{Origin server CPU usage}\label{fig:l1l2:cpu-server}
    \end{subfigure}
    \hfill
    \begin{subfigure}[b]{0.32\textwidth}
        \centering
        \includegraphics[width=0.95\textwidth]{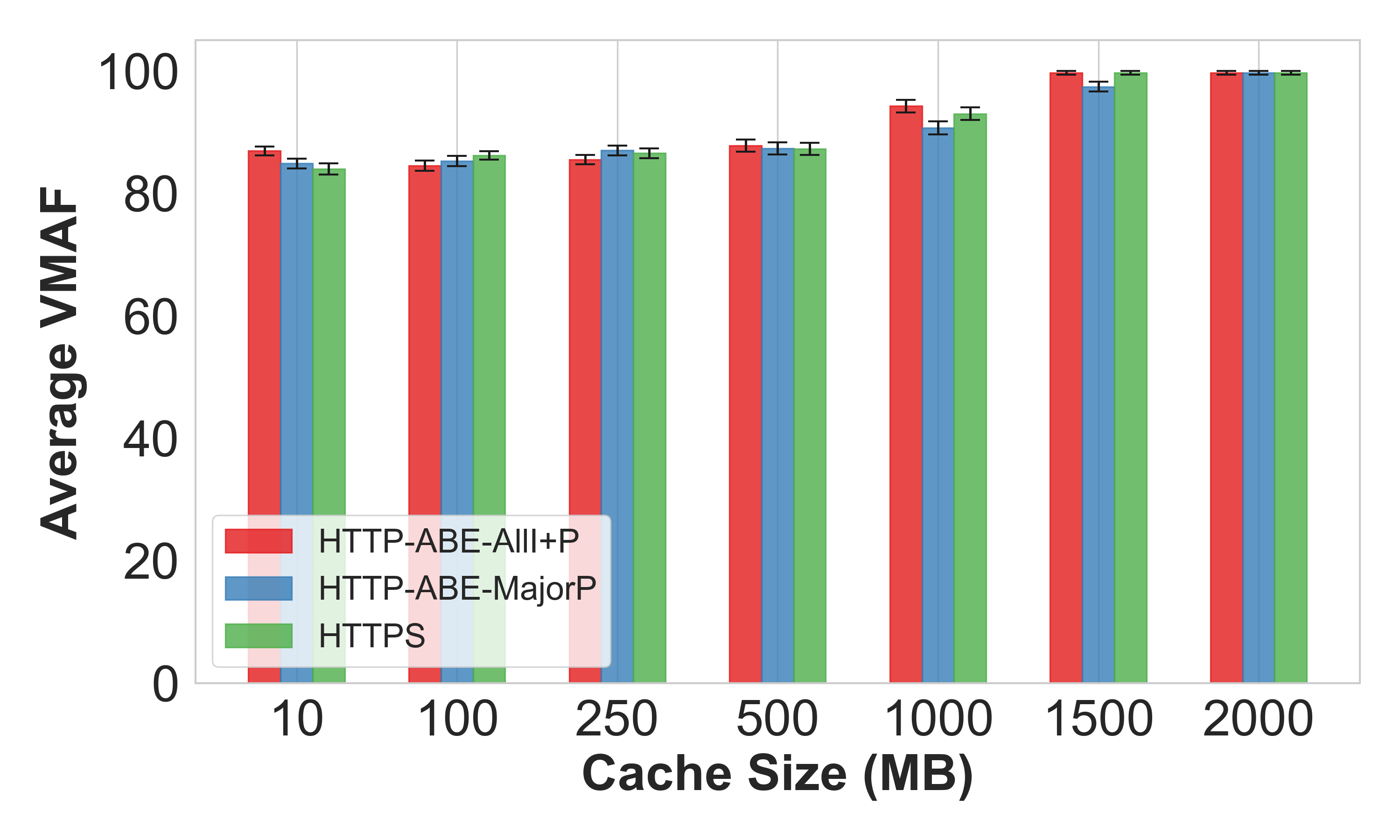}
        \caption{Average VMAF}\label{fig:l1l2:VMAF-abe-https}
    \end{subfigure}
    \hfill
    \begin{subfigure}[b]{0.32\textwidth}
        \centering
        \includegraphics[width=0.95\textwidth]{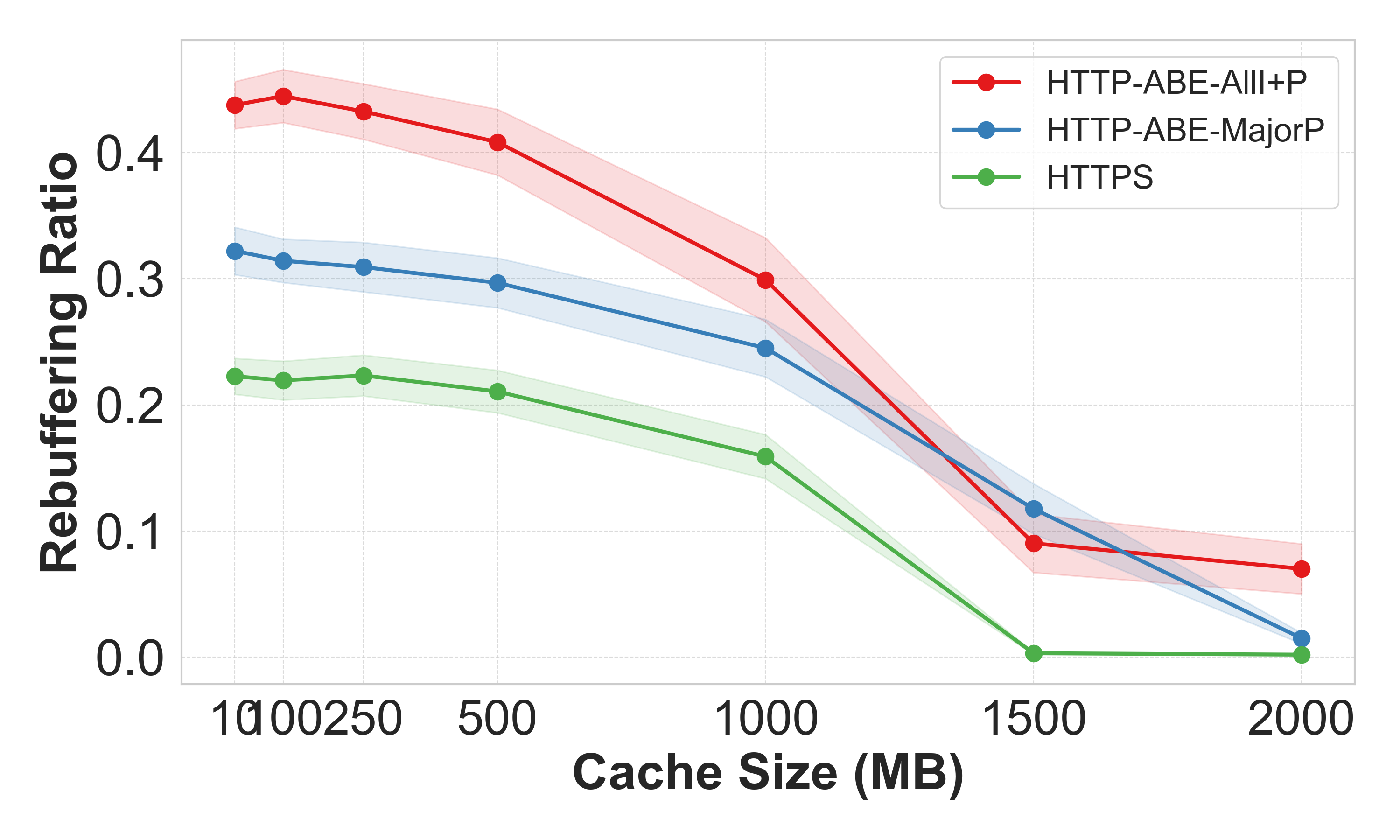}
        \caption{Rebuffering ratio}\label{fig:l1l2:rebuffer}
    \end{subfigure}

    \caption{Comparison of origin server CPU usage and client side QoE metrics for large-scale experiment.}
    \label{fig:l1l2:server-vmaf-rebuffer}
\end{figure*}

%\vspace{-0.1in}

\subsubsection{\textbf{Large Scale Hierarchical Experiment}} For the following evaluations, we focus on 2-second segments, which yielded more distinctive results in small-scale experiments.
\\
\label{subsubsec:large-scaleexp}
\textbf{CPU Load on L1 and L2 Caches:} Figure~\ref{fig:l1l2:cpu-cache} shows the average CPU usage for both L1 and L2 caches. Similar to the small-scale experiment, HTTPS consistently uses more CPU than both HTTP-ABE approaches. At the L1 cache, the CPU usage gap is most pronounced at smaller cache sizes. It narrows as cache size increases, eventually converging at 2000 MB, where most segments are served by the L2 caches, reducing the load on the L1 cache.

Between the two ABE schemes, ABE-allI+P continues to outperform MajorP in terms of CPU efficiency, as previously observed in the small-scale setup. The same trend applies to L2 caches: HTTPS consumes more CPU than both ABE schemes, and allI+P consistently uses less CPU than MajorP.

In case of the L1 cache, both ABE schemes consume up to 43\% less CPU than HTTPS at 10 MB. At 2000 MB, the reduction drops to 17\% for MajorP and 33\% for allI+P. For the L2 caches, MajorP achieves CPU savings ranging from 30–57\%, while allI+P performs even better, with reductions between 32–63\% compared to HTTPS.

These results differ from prior findings~\cite{10.1145/3712676.3714450}, where in large-scale experiments HTTP-ABE was observed to use more cache CPU than HTTPS at smaller cache sizes. In contrast, our results with \tsd video streaming consistently show lower CPU usage for both ABE approaches compared to HTTPS across all cache sizes. This discrepancy can be attributed to the nature of \tsd streaming, where four tiles are downloaded per segment, significantly increasing the load on the cache. Under HTTPS, this results in more frequent TLS termination, decryption, and re-encryption, contributing to higher CPU usage—an overhead avoided in the HTTP-ABE case.

\textbf{Comparison of Hit Rate:} Figure~\ref{fig:l1l2:hitrate} shows that hit rates for both HTTP-ABE approaches and HTTPS are comparable at the L1 cache. However, at the L2 caches, only ABE-MajorP exhibits lower hit rates between 1000 MB and 2000 MB—consistent with the trend observed in the small-scale experiment.

% \textbf{Comparison of Midgress Traffic:} As shown in Figure~\ref{fig:l1l2:midgress}, midgress traffic trends in the large-scale experiment closely mirror those observed in the small-scale setup. ABE-allI+P and HTTPS exhibit similar midgress levels across all three caches, while ABE-MajorP incurs noticeably higher midgress at larger cache sizes, consistent with earlier findings.

% \begin{figure*}[]\centering
%   \includegraphics[width=.95\linewidth]{figs/largescale-midgress_subplots.png}
%   \captionsetup{justification=centering}
%   \caption{Midgress traffic of L1 and L2 caches for different cache sizes}
%     \label{fig:l1l2:midgress}
%     %\vspace{-0.12in}
% \end{figure*}

\textbf{Server CPU, VMAF and Rebuffering:} Figure~\ref{fig:l1l2:server-vmaf-rebuffer} presents the comparison of server CPU usage~(\ref{fig:l1l2:cpu-server}), VMAF~(\ref{fig:l1l2:VMAF-abe-https}), and rebuffering~(\ref{fig:l1l2:rebuffer}) for the large-scale experiment. The observed trends closely align with those from the small-scale setup discussed in Sect.~\ref{subsubsec:smallscaleexp}, and can be interpreted using the same reasoning. 
%Therefore, we omit redundant explanation and refer the reader to the earlier discussion for detailed analysis.

% \vspace{-.15in}

% \begin{figure*}[!ht]\centering
%   % \includegraphics[width=3.5in]
%     \includegraphics[width=.95\linewidth]{newfigs/no-bw-cpu_usage_comparison_subplots.png}
%   \captionsetup{justification=centering}
%   \caption{CPU usage of L1 and L2 caches for different cache sizes without traffic congestion.}
%     \label{fig:no-bw-l1l2:cpu-cache}
% \end{figure*}
% % \textbf{Comparison of CPU load on origin Server without traffic congestion:}

%\input{5-related}
% \input{6-discussion}
%\vspace{-1em}
\section{Conclusions} \label{sec:conclusion}
This work presents a novel application of Attribute-Based Encryption (ABE) for HTTP-based \tsd video streaming
%, offering a secure and scalable alternative to traditional HTTPS. By 
where we leverage selective frame-based encryption to secure the content  rather than the transport layer. Our approach enables efficient caching and targeted access control without the overhead of TLS.

We implemented and evaluated two ABE-based selective encryption schemes: ABE-allI+P, a uniform encryption strategy applied across all tiles, and MajorP, a viewport-aware, two-dimensional encryption scheme that prioritizes frames in the viewer’s dominant field of view. 
%Through experiments on both small- and large-scale testbeds, 
We observe that HTTP-ABE significantly reduces cache CPU load up to 63\% and improved hit rates in some cases, outperforming HTTPS in both efficiency and scalability.  These benefits were especially pronounced in smaller segment durations, suggesting that ABE is well-suited for live streaming scenarios.

However, our evaluation also revealed important trade-offs. While both ABE schemes maintained comparable video quality (VMAF) to HTTPS, they incurred higher rebuffering. 

%ABE-allI+P suffereing, with ABE-allI+P suffering the most due to larger segment sizes and higher overhead. In contrast, 
We found MajorP offered a middle ground—with lower rebuffering than allI+P, smaller encryption and decryption overhead, and only slightly reduced content obfuscation. 
%Despite its weaker performance in terms of cache CPU usage and hit rate, 
MajorP demonstrates the potential of viewport-adaptive encryption strategies for balancing efficiency and security. We will explore such trade-offs and their implications on live-streaming in a future work.

\balance

\newpage

\bibliographystyle{ACM-Reference-Format}
\bibliography{refs, susmit}

\end{document}